\documentclass[12pt]{iopart}
\usepackage{graphics}
\usepackage{amssymb}
\begin{document}
\jl{3}

\title{Wedge covariance for 2D filling and wetting.}

\author{A. O. Parry, M. J. Greenall and A. J. Wood }
\address{Department of Mathematics, Imperial College 180 Queen's Gate, London SW7 2BZ, United Kingdom}

\begin{abstract}
  
A comprehensive theory of interfacial fluctuation effects occurring at 2D wedge
(corner) filling transitions in pure (thermal disorder) and impure (random
bond-disorder) systems is presented. 
Scaling theory and the explicit results of transfer matrix and replica trick studies of
interfacial Hamiltonian models reveal that, for almost all examples of
intermolecular forces, the critical behaviour at filling is fluctuation-dominated, characterised by universal critical exponents and scaling functions 
that depend only on the wandering exponent $\zeta$. Within this filling
fluctuation (FFL) regime, the critical behaviour of the mid-point interfacial
height, probability distribution function, local compressibility and
wedge free
energy are identical to corresponding quantities predicted for the
strong-fluctuation (SFL) regime for critical wetting transitions at planar walls.
In particular the wedge free energy is related to the SFL regime point tension
which is calculated for systems with random-bond disorder using the replica
trick.
The connection with the SFL regime for all these quantities can be expressed 
precisely in terms of special wedge covariance relations which complement standard 
scaling theory and restrict the allowed values of the critical exponents for
both FFL filling and SFL critical wetting. The predictions for the values of
the exponents in the SFL regime recover earlier results based on random-walk
arguments. The 
covariance of the wedge free energy
leads to a new, general relation for the SFL regime point tension which derives
the conjectured Indekeu-Robledo critical exponent relation and also explains 
the origin of the logarithmic singularity for pure systems known from exact 
Ising studies due to Abraham and co-workers. Wedge covariance is also
used to predict the numerical values of critical exponents and position
dependence of universal one-point functions for pure systems.
 \end{abstract}

\maketitle

\section{Introduction}

\begin{figure}[!h]
\begin{center}\resizebox{0.9\textwidth}{!}{%
  \includegraphics{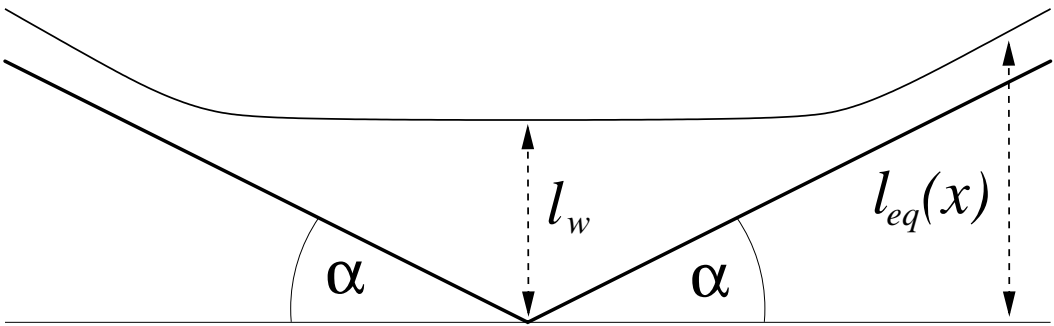}
}
\caption{Schematic portrait of the equilibrium interfacial height
$l_{eq} (x)$ for a fluid adsorbed in a wedge close to a filling
transition. $l_w$ denotes the mid-point interfacial height.}
\label{1wedge}
\end{center}
\end{figure}

   Fluids adsorbed near wedges, cones and corners show filling 
phenomena
\cite{Finn,Pomeau,Hauge,MFT,ourPRL1,ourJphysCM,ourPRL2,ourJPCMlett,ourJphysCM2,Nap,ourPRL3} similar to the wetting of planar wall-fluid and 
fluid-fluid interfaces
\cite{General,Forgacs}. Above a
filling temperature $T_{\mathrm{fill}}$ a wedge (say) in contact with vapour at bulk
coexistence is completely
filled by liquid, analogous to the complete wetting of a planar wall-fluid
interface above the wetting temperature $T_{\mathrm{wet}}$. There are, however a number of
notable distinctions between wetting and the different possible types of 
filling transition. Firstly, 
thermodynamic arguments \cite{Finn,Pomeau,Hauge} dictate that filling
precedes wetting, occurring when the contact angle satisfies $\theta=\alpha$,
where $\alpha$ is the wedge tilt angle (see \fref{1wedge}). Thus filling may occur 
in the absence of any wetting transition i.e. if the walls are partially wet up
to the bulk critical temperature. Secondly the conditions for observing 
continuous (critical) filling 
\cite{ourPRL2,ourJphysCM2} in the laboratory are much less restrictive
than for continuous (critical) wetting \cite{General}. For example 
3D cone or corner filling should be continuous provided the line-tension is 
negative. Finally, the critical singularities and 
fluctuation 
effects occurring at critical filling reflect the divergence of different 
length-scales compared to wetting. For example, in a 3D wedge, long-wavelength 
fluctuations in the 
interfacial height along the wedge dominate and lead to an  
interfacial roughness that is much larger than for wetting at a planar 
wall and which exhibits universal properties\cite{ourPRL2,ourJphysCM2,Nap}. 
 
For 2D wedges (and 3D cones) on the other hand
something quite different and unexpected occurs. Recent studies based on 
effective interfacial Hamiltonians 
\cite{ourPRL1,ourJphysCM,ourJphysCM2} and more microscopic 
Ising models \cite{ourPRL3}  indicate that there is a
fundamental connection with the strong-fluctuation (SFL) regime of critical 
wetting \cite{Forgacs,Lipowsky,Fisher,LandF,Wuttke}. More specifically, even in the 
presence of long-ranged forces, the 
divergence of the interfacial height at critical filling together with the precise scaling
form of the mid-point interfacial height probability distribution function
(PDF) are identical to predictions for the critical wetting SFL regime. 
In other words the substrate geometry effectively turns off the 
influence of long-ranged solid-fluid and fluid-fluid forces so the fluid mimics
fluctuation behaviour predicted for planar systems with short-ranged forces. 
 In 2D this has been demonstrated for both pure and impure systems 
(with random-bond disorder) and can be precisely expressed in terms of special
wedge
covariance laws which relate the interfacial heights and 
PDF's in the different geometries at bulk two-phase coexistence 
\cite{ourJphysCM,ourJphysCM2}. Whilst the
demonstration of these laws, particularly for impure systems, is rather 
technical, the final expressions are extremely simple and contain a great deal
of information.  Consider a fluctuation-dominated filling transition occurring in a 2D wedge and let $l_w(\theta,\alpha)$ 
and $P_w(l;\theta,\alpha)$ 
denote the equilibrium mid-point interfacial height and corresponding PDF respectively. Now let 
$l_{\pi}(\theta)$ and $P_{\pi}(l;\theta)$ denote the interfacial height and 
PDF for a SFL regime wetting transition, occurring at $\it{planar}$ wall-fluid 
interface, written in terms of the contact angle. Covariance for the interfacial
heights implies that at bulk coexistence and as $\theta\to\alpha$,
\begin{equation}
l_w(\theta,\alpha)=l_\pi(\theta-\alpha)
\label{law01}
\end{equation}
The statement of covariance for the PDF's is even stronger:
\begin{equation}
P_w(l;\theta,\alpha)=P_{\pi}(l;\theta-\alpha)
\label{law02}
\end{equation}
implying that not only the interfacial height but also the roughness (and all 
moments of the distribution) at filling and wetting are similarly related.

 In this paper we further investigate the connection between filling and
wetting and demonstrate covariance relations for other
quantities of interest. Again, 
because the derivation of these results is rather
technical we quote their final form here. For the excess free energy of the 
wedge $f_w(\theta,\alpha)$ at coexistence we find that for both pure
 and impure systems
\begin{equation}
f_w(\theta,\alpha)=\tau(\theta)-\tau(\theta-\alpha)
\label{law03}
\end{equation}
where $\tau(\theta)$ denotes the point tension near a  
SFL regime critical wetting transition. We also consider the local 
compressibility for filling
$\chi_w(z;\theta,\alpha)$ corresponding to the derivative of the mid-point density
profile w.r.t. chemical potential $\mu$ evaluated at two-phase coexistence.
This is related to the corresponding expression $\chi_{\pi}(z;\theta)$
for SFL regime wetting by
\begin{equation}
\chi_w(z;\theta,\alpha)=\left(\frac{\theta}{\alpha} - 1 \right)\chi_{\pi}(z;\theta-\alpha)
\label{law04}
\end{equation}
  
 These covariance relations complement the standard scaling 
hypothesis for the singular contribution to the corner free-energy and may be
understood heuristically by considering the special influence that the wedge 
(and also cone) geometry has on interfacial configurations. The relations are
 rather restrictive and from them we may deduce the allowed 
values of the critical exponents for both 2D filling $\it{and}$ critical wetting 
transitions without explicit model calculations. This is somewhat analogous to
the restrictions imposed by the principle of conformal invariance for bulk
and surface critical phenomena \cite {BPZ}. We will also use the covariance
relations to derive new results for the point tension and 
position dependence of the local compressibility. Indeed for the point tension 
we will be able to derive a conjectured critical exponent relation due to Indekeu and Robledo
\cite{IR,I} and also explain why, for pure systems, the 
point tension shows a logarithmic singularity as $\theta\to 0$ 
\cite {ABU}.

Our article is arranged as follows: In \S II we present the necessary background 
theory for critical wetting describing, in turn, the derivation of critical
exponent relations and fluctuation regimes from heuristic scaling theory and
the scaling of the density profile,
probability distribution function and their short-distance expansions. In 
\S III we discuss the singularities of the point tension first recalling the hyper-scaling 
conjecture of Indekeu and Robledo \cite{IR} and exact
Ising model results for pure systems due to Abraham,
Latr\'emoli\`ere and Upton 
\cite{ABU}. We then present a lengthy calculation of the point tension $\tau$ for 
critical wetting in pure and impure systems systems using continuum interfacial
models.  As far as we are aware $\tau$ has not been calculated before for
systems with random-bond disorder and the derived 
expression shows singular behaviour as $\theta\to 0$ in full agreement with 
the Indekeu-Robledo conjecture. In \S IV we begin our discussion of 2D filling 
transitions and first discuss critical exponent relations, heuristic
fluctuation theory and the scaling of the density profile and PDF, which 
parallels our earlier discussion of critical wetting. In \S V we recall the 
main results of explicit effective Hamiltonian studies of filling which will 
be further developed here. These calculations support the scaling theory
developed in \S IV and also demonstrate the covariance laws (\ref{law01})-(\ref{law03}) quoted above. The consequences of these relations and the
restrictions they place on the allowed values of the critical exponents at 2D
filling and wetting (in systems with short-ranged forces) are discussed
at length. A new and very simple result for the point tension is also discussed
from which we can derive the Indekeu-Robledo exponent relation. 
In \S VI we turn our attention to scaling behaviour occurring off 
bulk coexistence and focus on the position dependence of the local 
compressibility leading to the final covariance relation (\ref{law04}). Finally
we revisit the nature of filling and wetting in pure systems and use the
covariance relations to re-derive the specific results of the transfer matrix
studies. We 
finish our article with a brief summary of our main results and discussion of
future work.

\section{2D critical wetting and SFL regime}  
 
 We begin with a brief survey of the central results from the theory of
critical wetting paying special attention to critical exponents, fluctuation 
regimes, and the scaling of the density profile and PDF. Further
details and original references can be found in the excellent review 
articles \cite{General,Forgacs}. 

\subsection{Critical exponents and exponent relations}

Consider the interface between a planar substrate (wall), located
in the $z=0$ plane, and a 
bulk fluid phase at temperature $T$ and chemical potential 
$\mu$ (and corresponding pressure $p$). The two-phase coexistence line is
denoted $\mu=\mu_{\mathrm{sat}}(T)$ with corresponding bulk liquid and vapour phases 
densities, $\rho_l$ and $\rho_v$ respectively. Throughout this article we set $k_B
T\equiv 1$ for convenience. We suppose the 
wall-vapour interface preferentially adsorbs 
the liquid phase and is completely wet above a wetting transition 
temperature $T_{\mathrm{wet}}$. Thus the wetting phase boundary is defined by the
vanishing of the contact angle
\begin{equation}
\theta(T)=0; \qquad T\ge T_{\mathrm{wet}}
\label{wetting}
\end{equation}
 The wetting transition corresponds to a singularity in the excess free-energy (surface tension) of the planar 
wall-vapour interface $\sigma_{wv}$. This is defined \cite{Row} by subtracting
the bulk contribution from the total grand potential $\Omega$ 
\begin{equation}
\sigma_{wv}=\frac{\Omega+pV}{A}
\label{sigma}
\end{equation} 
in the limit of infinite volume V and planar area A. On approaching a critical wetting
transition the 
adsorption $\Gamma\approx(\rho_l-\rho_v)l_\pi$ with 
$l_\pi$ the equilibrium height of the liquid-vapour
interface above the wall, diverges continuously. The critical wetting transition
has two relevant scaling fields which we can identify (for fixed strength of
the wall-fluid intermolecular potential) as $t' = (T_{\mathrm{wet}}
-T)/T_{\mathrm{wet}}$ and $h =
(\rho_l-\rho_v)(\mu_{\mathrm{sat}}- \mu)$. From scaling theory we anticipate that off 
two phase coexistence $\sigma_{wv}$ contains a singular term, $f_{\mathrm{sing}}$  
which vanishes at the critical wetting phase boundary and can be written
\begin{equation}  
 f_{\mathrm{sing}}= t'^{2-\alpha_s} W_{\pi}(ht'^{-\Delta})
\label{freeenergywetting}
\end{equation}
where $\alpha_s$, $\Delta$ are the surface specific heat and gap exponents
respectively, $W_{\pi}(x)$ is the scaling function and we have restricted our attention to $t'\ge 0$. Now at coexistence
we have, by definition $f_{\mathrm{sing}}\equiv
\sigma_{wv}-(\sigma_{wl}+\sigma_{lv})$ so that from Young's
equation \cite{Row} it follows that the contact angle vanishes as
\begin{equation}
 \theta \sim t'^{(2-\alpha_s)/2}
\label{thetasing}
\end{equation}
and can be used as an alternative measure of the temperature-like scaling 
field. In addition to the mean interface height $l_\pi$ we also need consider the divergence of the r.m.s interfacial width or
roughness $\xi_{\perp}$ and the transverse correlation length
$\xi_{\parallel}$ measuring the correlations in fluctuations of the 
interfacial height along the wall. At $h=0$, the divergence of
these length-scales is characterised by critical exponents defined by
\begin{equation}
l_\pi\sim t'^{-\beta_s},\xi_{\perp} \sim t'^{-\nu_{\perp}},
\xi_{\parallel} \sim t'^{- \nu_{\parallel}} 
\label{length}
\end{equation}
and we expect scaling behaviour
similar to (\ref{freeenergywetting}) off coexistence. Thus, along the 
wetting critical isotherm ($T=T_{\mathrm{wet}}$, $h\to 0$) we define
\begin{equation} 
l_\pi\sim h^{-\psi}
\label{lh}
\end{equation}
 and anticipate that the critical exponent 
$\psi=\beta_s/\Delta$. Critical exponent
relations immediately follow from the scaling hypothesis. Firstly from the 
Gibbs adsorption equation \cite{General,Row}
\begin{equation}
\partial f_{\mathrm{sing}}/\partial h=l_\pi
\label{Gibbs}
\end{equation}
we have
\begin{equation}
\Delta=2-\alpha_s+\beta_s
\label{del}
\end{equation}
 which identifies the gap exponent. Secondly, from the compressibility sum-rule
(see for example \cite{General})
\begin{equation}
\partial^2 f_{\mathrm{sing}}/\partial h^2 \propto \xi_{\parallel}^2
\label{comp}
\end{equation}
we have $2-\alpha_s-2\Delta=-2 \nu_{\parallel}$ so that on eliminating the gap exponent
we arrive at the important relation
\begin{equation}
2-\alpha_s=2\nu_{\parallel}-2\beta_{s}
\label{Rush}
\end{equation}
which is valid for all dimensions, ranges of forces and fluctuation regimes of
interest.
 The perpendicular and transverse correlation lengths are related through
the wandering exponent defined by \cite{Fisher}
\begin{equation}
 \xi_{\perp} \sim \xi_{\parallel}^\zeta
\label{wandering}
\end{equation}
with the value of $\zeta\ge 0$  dependent on the dimensionality and qualitative type of 
disorder. For discussions of wetting and filling in 2D systems the most relevant
values are $\zeta=1/2$ and $\zeta=2/3$ for pure (thermal disorder) and impure
 systems (random-bond disorder) respectively. Recall that random-fields
destroy phase coexistence in two dimensions so cannot be considered. Also
whilst values of the wandering exponent $\zeta<1/2$ have been predicted for
some models of interfacial roughening transitions in quasi-crystalline materials
this is not of particular importance to the general fluctuation theory of
wetting (and filling) and will not be considered here. The predictions we make
for the values of critical exponents at filling will assume that in 2D, 
$1>\zeta\ge 1/2$.

\subsection{Fluctuation regimes for wetting} 

  Quite generally, provided $\zeta > 0$, the critical singularities at wetting 
are believed to fall into $\it{three}$ distinct classes depending on the interplay 
between interfacial wandering and the `direct' influence of
intermolecular forces \cite{Forgacs,Lipowsky,Fisher,LandF,Wuttke}. The latter 
can be modelled by the binding potential  
\begin{equation}
W(l) = -\frac{a}{l^p} + \frac{b}{l^q} + \dots ; l > 0 
\label{BPot} 
\end{equation}
with $a,b$ effective Hamaker constants and indices $q>p>0$ depending on the range
of the intermolecular forces. The binding potential describes the bare or mean-field wetting
transition and in order for this to be continuous we require that $a=0,b>0$ 
 at the (mean-field) phase boundary. Thus $a\propto
(T_{\mathrm{wet}}^{MF}-T)$  where $T_{\mathrm{wet}}^{MF}$ is the mean-field wetting temperature. The 
existence of three regimes can be understood semi-quantitatively \cite{LandF} 
by comparing the bare binding potential with an effective fluctuation
contribution
\begin{equation}
W_{fl}(l)\approx \frac{\Sigma\xi_{\perp}^2}{2\xi_{\parallel}^2}
\label{landf}
\end{equation}
the form of which is suggested by interfacial Hamiltonian models. Beyond
mean-field level we anticipate large scale fluctuation effects with
$l_{\pi}\sim\xi_{\perp}$ and thus using (\ref{wandering}) one can estimate
\begin{equation}
\frac{\xi_{\perp}}{\xi_{\parallel}}\approx l^{1-1/\zeta}
\label{precursor}
\end{equation}
implying that $W_{fl}(l)\approx l^{2(1-1/\zeta)}$. The competition between
$W_{fl}(l)$ and the bare potential leads to the following three fluctuation 
regimes
\begin{itemize}
\item{{\bf Mean field (MF) regime} - If $q<2(1/\zeta-1)$ fluctuation
effects are negligible, $l_{\pi}\gg\xi_{\perp}$ and the mean-interface position remains close to
the minimum of the binding potential. The phase boundary remains $a=0$ with 
$\beta_s=1/(q-p)$.}
\item{{\bf Weak fluctuation (WFL) regime} - If $q>2(1/\zeta-1)$ but $p<2(1/\zeta-1)$
the repulsion from the wall has an entropic origin but the attraction is still
due to long-ranged forces. In this regime the phase boundary remains $a=0$ with
\begin{equation}
\beta_s=\frac{1}{2(1/\zeta-1)-p}
\label{WFLB}
\end{equation}
with large-scale interfacial fluctuations $l_{\pi}\sim\xi_{\perp}$.}
\item{{\bf Strong fluctuation (SFL) regime} - If $p>2(1/\zeta-1)$, fluctuations
dominate leading to a renormalisation of the phase boundary and universal
critical behaviour. Because the wetting phase boundary no longer occurs at
$a=0$ one cannot use the above heuristic argument to determine the
values of the critical exponents. However a remarkable feature of
2D wetting is that the values of the SFL regime critical exponents can 
be explicitly related to the wandering exponent $\zeta$ using very general
random walk arguments \cite{Fisher}. For wetting in systems with 
$\zeta \ge 1/2$ the full set of values for the critical exponents is 
\begin{equation}
\alpha_s=0,\beta_s=\frac{\zeta}{1-\zeta},
\nu_{\parallel}=\frac{1}{1-\zeta}
\label{SFL1}
\end{equation}
and
\begin{equation}
\Delta=\frac{2-\zeta}{1-\zeta},\psi=\frac{\zeta}{2-\zeta}
\label{SFL2}
\end{equation}
which are in precise agreement with the explicit results of Ising model
\cite{Abraham} and
effective interfacial Hamiltonian studies
\cite{Burkhardt1,Chalker,Kardar,Burkhardt2}. The values of these critical
exponents will play a central role in our discussion of 2D filling.} 
\end{itemize}

Finally we point out that
in both the WFL and SFL regime, where $l_{\pi}\sim\xi_{\perp}$ the
divergence of the interfacial height, at $h=0$, written in terms of the contact
angle,
\begin{equation}
l_{\pi}(\theta)\sim \theta^{-\hat{\beta_s}}
\label{hat}
\end{equation}
is characterised by a universal critical exponent
\begin{equation}
\hat{\beta_s}\equiv\frac{2\beta_s}{2-\alpha_s}=\frac{\zeta}{1-\zeta}
\label{weak}
\end{equation}
which follows directly from the critical exponent  relation
(\ref{Rush}) $\it{without}$ using the explicit values of the critical exponents
(\ref{WFLB}),(\ref{SFL1}). We shall return
to this later when we use the covariance relations to determine the values of
the critical exponents at 2D filling.

\subsection{Scaling of the PDF and the short-distance expansion}  

The position dependence of the equilibrium density profile, $\rho(z)$, local response functions such as 
the compressibility/susceptibility
$\chi(z)\propto\partial\rho(z)/\partial \mu$ and also higher-point functions all show
scaling behaviour in the WFL and SFL scaling regimes. The scaling emerges
 in the appropriate 
limits
$z\to \infty,t'\to 0,h \to 0$ with $zt'^{\beta_s},ht'^{-\Delta}$ arbitrary and
for the profile we anticipate \cite{Parry,Parry2,Parry3}
\begin{equation}
\rho(z)=\rho_l-(\rho_l-\rho_v)\Xi_{\pi}(zt'^{\beta_s},ht'^{-\Delta})
\label{rhoscalewet}
\end{equation}
where $\Xi_{\pi}(x,y)$ is the scaling function satisfying
$\Xi_{\pi}(0,y)=0$, $\Xi_{\pi}(\infty,y)=1$ and which is distinct in the SFL and WFL regimes. Clearly the scaling 
of $\rho(z)$ does not include oscillatory structure close to the wall 
or effects associated with bulk and surface criticality but rather reflects 
the large scale fluctuations of the unbinding liquid-vapour
interface. For large $z$ but $zt'^{\beta_s}\to 0$ the profile has a
characteristic algebraic short-distance expansion (SDE)
\cite{Parry,Parry2,Parry3}
\begin{equation}
\rho(z)-\rho_l \approx (\rho_v-\rho_l)(zt'^{-\beta_s})^\gamma
\label{gamma}
\end{equation}
where, for simplicity we have set $h=0$. The SDE critical exponent $\gamma$
(referred to as $\theta$ in earlier work) also describes the behaviour of the local
compressibility and pair-correlation functions close to the wall. The critical
exponent $\gamma$ is not independent, and can be related to others using
standard surface Maxwell conditions and sum-rules. 
Importantly it takes different universal values in the SFL and WFL regimes and
can be identified as \cite{Parry2,Parry3}
\begin{equation}
\gamma^{SFL}=2(1/\zeta-1)-1/\beta_s
\label{SDE2}
\end{equation}
and
\begin{equation} 
\gamma^{WFL}=2/\zeta-1
\label{SDE3}
\end{equation}
which are valid for arbitrary dimensions. The scaling of the profile is directly related to 
the scaling of the interfacial height PDF, $P_{\pi}(l)$, since interfacial fluctuations
dominate the distribution of matter and we may write
\begin{equation}
\rho(z)=\rho_l-(\rho_l-\rho_v)\int_{0}^{z}P_{\pi}(l) dl
\label{density}
\end{equation}
where we have assumed the interface separates regions of bulk vapour and 
liquid density and we have omitted the field dependence of the PDF. Throughout
 this paper we shall omit the field dependence whenever the 
equation containing it is exact within effective Hamiltonian theory and not 
just restricted to the asymptotic critical regime. Since the contact angle $\theta$ is an equivalent (possibly non-linear) 
measure of the temperature-like scaling field $t'$ it is possible to  write the
scaling dependence as $P_{\pi}(l;\theta,h)$ rather than 
$P_{\pi}(l;t',h)$. Moreover for the next few sections we concentrate on
behaviour occurring at $h=0$ and define 
$P_{\pi}(l;\theta,0)\equiv P_{\pi}(l;\theta)$. Scaling then implies that in 
the WFL and SFL regimes
\begin{equation}
P_{\pi}(l;\theta)=\tilde{a}\theta^{\hat{\beta_s}}\Lambda_{\pi}(\tilde{a}l\theta^{\hat{\beta_s}})
\label{Pdefscaling}
\end{equation}
where $\Lambda_{\pi}(x)$ is the scaling function and $\tilde{a}$ is
a suitable metric factor having dimensions of inverse length. This may be
chosen so that the
 argument of the scaling function is equivalent to the length-scales ratio
$l/l_\pi(\theta)$. Notice that the power-law dependence of the contact angle 
follows from (\ref{weak}) and is the same in the WFL and SFL regimes. The
different fluctuation effects occurring in these regimes are distinguished by 
the appropriate scaling functions $\Lambda_{\pi}^{SFL}(x)$ and 
$\Lambda_{\pi}^{WFL}(x)$. Both functions are
similar at large distances where, out of the range of the binding potential,
 they decay exponentially quickly but have quite
distinct short-distance expansions $\Lambda_{\pi}(x)\sim x^{\gamma-1}$
describing the limit $l/l_\pi\to 0$. Using the appropriate values for the critical 
exponents in 2D it can be seen that, for both pure and impure systems, the interface makes many more 
excursions to the wall in the SFL regime than in the WFL regime. Hereafter
we will only need to deal with the properties of the SFL regime. 

The explicit results of effective interfacial Hamiltonian studies
are completely consistent with the scaling predictions and SDE. For pure
systems with just thermal disorder the PDF in the SFL regime is 
particularly simple \cite {ourJphysCM,Burkhardt2}
\begin{equation}
P_{\pi}(l;\theta)=2\Sigma\theta e^{-2\Sigma\theta l}
\label{PDFThermal}
\end{equation}
where $\Sigma$ denotes the stiffness coefficient of the unbinding 
interface and may be identified with $\sigma_{lv}$ for continuum, fluid-like
systems for which the interface is isotropic. For random bond disorder the 
expression for
$P_{\pi}(l;\theta)$ is considerably more complicated but can still be 
calculated analytically using replica trick methods\cite{Forgacs} 
\begin{equation}
P_{\pi}(l;\theta)=\frac{\Sigma\theta}{\pi\sqrt{2l\kappa}}e^{-l\theta^2\Sigma^2/2\kappa}
\int_{0}^{\infty}ds\frac{\sqrt{s}e^{-s/4}}{s+2l\theta^2\Sigma^2/\kappa}
\label{PDFRB}
\end{equation}
where $\kappa$ is the inverse length-scale associated with the disorder (see
later). Note that in both these expressions the respective 
combinations $l\theta$ and 
$l\theta^2$, together with the SDE's, are in agreement with
the above scaling theory.

\section{The point tension for 2D wetting in pure and impure systems}

\subsection{The Indekeu-Robledo conjecture}

 The one ingredient missing in our review of fluctuation effects at 2D wetting
is the nature of the point tension $\tau$ measuring the excess
free-energy associated with the point of three-phase contact between wall-vapour and
wall-liquid interfaces \cite{IR,I,ABU,DB,Steph}. The reason for this, as first 
pointed out by Abraham, Latr\'emoli\`ere and Upton (ALU)\cite{ABU}, is that beyond mean-field level,
fluctuation effects make the definition of $\tau$ a rather subtle issue. The
purpose of this long section is to identify a method of defining $\tau$ within
continuum effective interfacial Hamiltonian theory that we can apply to the case of
wetting with random-bond disorder. To begin recall that within mean-field
theory, it is straight-forward to to define the point/line tension $\tau$ by
simply subtracting the necessary bulk and interfacial contributions from the 
grand-potential of the heterogeneous wall-fluid interface \cite{IR,I,DB}. Such 
studies reveal that as $T\to T_{\mathrm{wet}}$ at $h=0$, $\tau$ contains a singular 
contribution which we write
\begin{equation}
\tau_{\mathrm{sing}}\sim t'^{2-\alpha_l}
\label{line}
\end{equation}
with a point/line tension specific heat exponent $\alpha_l$ which depends sensitively
on the range of the forces. A crucial insight into the singularities of the
point/line tension was made by Indekeu and Robledo \cite{IR} who pointed out that within
mean-field theory the singularities of $\tau$ were consistent with the 
hyper-scaling equation 
\begin{equation}
\alpha_l=\alpha_s+\nu_{\parallel}
\label{Joseph1}
\end{equation}
and conjectured that this is generally valid even in the presence of
 fluctuation effects. The Indekeu-Robledo conjecture is important
because it relates the excess free-energy of a heterogeneous 
wall-fluid interface to the properties of a homogeneous wall-vapour interface.
As we shall show there are very good reasons for regarding this as the
precursor of a covariance relationship between filling and critical wetting. 
Indeed, we shall be able to derive a precise relation for the
point tension, valid for pure and impure systems, which is in perfect agreement
with (\ref{Joseph1}). Assuming the validity of the hyperscaling 
relation for the point tension in 2D we can identify, in the SFL regime 
\begin{equation}
2-\alpha_l=\frac{1-2\zeta}{1-\zeta}
\label{alphal}
\end{equation}
which completes our list of critical wetting exponents. 

 Following our earlier discussion of the interfacial height and PDF it is
convenient for later purposes to measure the singular contribution to the 
point tension as a function of the contact angle rather than $t'$. We will only
consider the point tension for systems with short-ranged forces (belonging to
the SFL regime) and thus expect
\begin{equation}
\tau_{\mathrm{sing}}(\theta)\sim\theta^{2-\alpha_l}
\label{TauSFL}
\end{equation}
with $2-\alpha_l$ given by (\ref{alphal}) In fact, as we shall show, for interfacial models with strictly short-ranged 
(contact)
forces there is no ambiguity defining the point tension itself $\tau(\theta)$
to be a function of $\theta$, although, of course such models are only
well-defined when $\theta$ is small. Thus, for pure systems 
with $\zeta=1/2$, the Indekeu-Robledo hyper-scaling relation predicts 
$2-\alpha_l=0$  which may either 
mean that $\tau$ remains finite at $T_w$ or diverges (or vanishes) more
slowly than
 any power law. For random bonds however, the Indekeu-Robledo 
prediction is unambiguous: $\alpha_l=3$ so that $\tau_{\mathrm{sing}}\sim\theta^{-1}$. 
These predictions are discussed in detail below. 

\subsection{The ALU point tension for pure systems}

ALU point out that, at least for two-dimensional systems, considerable
care has to be taken in 
defining the point tension due to the influence of large scale interfacial 
wandering which smooths out the point of contact. In mean-field
theories, which ignore fluctuation effects, there is no pathology
involved in constructing 
boundary conditions which induce a line of contact between a wall-vapour interface and 
the edge of an infinite drop. However, this latter concept becomes ill-defined as
soon as fluctuation effects are introduced since the surface of an infinite drop of liquid has unbounded fluctuations in the interfacial height. To overcome this 
problem ALU propose a fluctuation theory-based definition of $\tau$,
 involving a convolution of partition functions. This requires as input
 some appropriate choice for the partition function representing a 
finite-size liquid drop. In this way the thermodynamic limit can be
taken yielding a well-defined point tension, although the expression obtained
depends crucially on the choice of restricted partition function used 
to model the edge of the liquid drop. The most 
satisfactory definition shows the singular behaviour 
\begin{equation}
\tau_{\mathrm{sing}}(\theta)\sim -\ln\theta
\label{line2D}
\end{equation}
close to the wetting transition which is indeed consistent with the Indekeu-Robledo 
exponent relation. The other fluctuation definitions considered 
by ALU yield point tensions that are either non-singular or have 
a different numerical pre-factor of the logarithm (other than unity). However as 
pointed out by ALU the above divergence is appealing since it coincides precisely with the singularity predicted by a heuristic 
energy-entropy balance argument. In an infinite Ising strip of width $L$ 
lattice spacings and with opposing surface fields (which we refer to as a 
$\bf{+-}$ strip), it is well understood that pseudo-phase coexistence only occurs 
below the wetting temperature and for sufficiently large strip widths
\cite{Binder,Parry and Evans1,Parry and Evans2,AU}. This behaviour
is characterised by an exponentially large correlation length (see their Fig. 3)
reflecting the asymptotic degeneracy of the lowest two transfer-matrix 
eigenvalues. Physically this mean that the interface sticks to each wall over
exponentially large distances which can be estimated by
\begin{equation}
\frac{\xi_{\parallel}}{\xi_0}\sim e^{\Sigma\theta L+2\tau} 
\label{expo}
\end{equation}
valid for large $L$, small $\theta$ and $\theta L\to\infty$. Here
$\xi_0$
is an appropriate (non-singular) length-scale for measuring distances along the
strip which we anticipate is of the order of the 
bulk correlation length. This has to be introduced for dimensional reasons and 
plays no role in determining the divergence of $\tau$ near wetting. The 
argument of the exponential
reflects the free-energy cost of an interface jump from one wall to the other
with contributions arising from surface free-energies, leading to the  
$\Sigma\theta L$ term and two point tensions. Exact evaluation of the 
correlation length in the $\bf{+-}$ Ising model and also in solid-on-solid 
approximation (valid at low temperatures away from the
bulk critical point) yields a point tension in precise agreement with result
(\ref{line2D}) from the ALU convolution definition. 

\subsection{Continuum interfacial models of the point tension in pure systems}

The purpose of this subsection is to show that one may also obtain the
logarithmic singularity of the point tension for pure systems from the
asymptotic scaling of the PDF as evaluated using a continuum effective
interfacial Hamiltonian. The advantage of this approach is that it can be
readily generalised to systems with random-bond disorder, which we will
consider next. To begin, we introduce the interfacial model and explicitly 
evaluate the point tension using the ALU finite-size $\bf{+-}$ strip 
identification discussed above.

The general fluctuation theory of wetting 2D in pure systems, without
quenched impurities, is based on the interfacial model (see \cite{Forgacs} and
references therein)
 \begin{equation} 
H[l] = \int dx \left\{ \frac{\Sigma}{2} \left(\frac {dl}{dx} \right)^2 + hl+W(l) \right\}
\label{effham}
\end{equation}
where $l(x)$ is the local height of the unbinding (liquid-vapour) at position
$x$ along the wall and $W(l)$, $\Sigma$ are the binding potential and stiffness
coefficient introduced earlier. We emphasise again that we will focus on isotropic 
bulk fluid systems and identify $\Sigma$ with the surface tension 
$\sigma_{lv}$. The partition function $Z_\pi(l_1,l_2;X)$ of a interface of length $X$ with
fixed end positions $l(0)=l$,$l(X)=l'$ is expressed in spectral form using
continuum transfer-matrix methods \cite{Burkhardt2},
\begin{equation}
Z_\pi(l,l';X)=\sum_n \psi^*_n(l')\psi_n(l)e^{-E_nX}
\label{prop}
\end{equation}
where the eigenvalues and eigenfunctions, labelled $n=0,1,2...$ satisfy the Schr\"odinger equation
\begin{equation}
-\frac{1}{2\Sigma}\frac{d^2\psi_n}{dl^2}+(hl+W(l))\psi_n(l)=E_n\psi_n(l) 
\label{Schro}
\end{equation}
Thus in the thermodynamic limit $X\to\infty$ the singular part to the 
free-energy is simply 
\begin{equation}
f_{\mathrm{sing}}=E_0
\label{fE}
\end{equation}
which, for $h=0$ and $T<T_w$ allows us to identify the contact angle
\begin{equation}
\theta=\left(\frac{-2E_0}{\Sigma}\right)^{1/2}
\label{t2D}
\end{equation}
Here we have used Young's equation in the small contact limit for which the
interfacial model is valid. Similarly the normalised interfacial height PDF and parallel correlation length 
follows as
\begin{equation}
P_\pi(l)=\vert \psi_0(l)\vert^2
\label{Born}
\end{equation}
and
\begin{equation}
\xi_{\parallel}=\frac{1}{E_1-E_0}
\label{xi}
\end{equation}
respectively. For future reference we also define the matrix elements
\begin{equation}
\langle m\vert f(l) \vert n \rangle=\int dl \psi_m^*(l) f(l)\psi_n(l)
\label{matrix}
\end{equation}
which will appear in the transfer matrix theory for the wedge geometry.
  As discussed by Burkhardt \cite{Burkhardt2}, the scaling form of the PDF at
(\ref{PDFThermal}) at $h=0$ 
characteristic of the SFL
regime, together with the values of the critical exponents quoted earlier (with
$\zeta=1/2$), readily emerges from the transfer matrix formulism if, instead of 
the binding potential contribution to (\ref{effham}), one imposes the boundary
condition on the wave-functions
\begin{equation}
\psi_n'(0)=-\lambda\psi_n(0)
\label{bound1}
\end{equation}
where, for pure systems, $\lambda\propto t'$. Specifically, the ground-state 
energy $E_0= -\lambda^2/(2\Sigma)$, contact angle $\theta=\lambda/(\Sigma)$ and mean interfacial height
$l_\pi=1/(2\Sigma\theta)$. 

Next consider the interfacial model of the finite width $\bf{+-}$ Ising strip
at bulk coexistence.
Since the system has short-ranged forces we can mimic the influence of the
surface fields through the boundary conditions
\begin{equation}
\psi_n'(0)=-\lambda\psi_n(0),\psi_n'(L)=\lambda\psi_n(L)
\label{b2}
\end{equation}
For $T_{\mathrm{wet}}>T$ and large $L$ the first two eigenfunctions are
\begin{equation}
\psi_0(l)\propto\cosh(\sqrt(2\Sigma \vert E_0\vert)(l-L/2))
\label{+-eigf1}
\end{equation}
and
\begin{equation}
\psi_1(l)\propto\sinh(\sqrt(2\Sigma \vert E_1\vert)(l-L/2))
\label{+-eigf2}
\end{equation}
 Using the boundary conditions (\ref{b2}) we find for $\lambda L\to\infty$
\begin{equation}
E_0=-\frac{\lambda^2}{2\Sigma}(1+4e^{-\lambda L}+...)
\label{+-eigv1}
\end{equation}
and
\begin{equation}
E_1=-\frac{\lambda^2}{2\Sigma}(1-4e^{-\lambda L}+...)
\label{+-eigv2}
\end{equation}
Writing these in terms of the contact angle of the semi-infinite geometry we
arrive at the desired expression for the parallel correlation length 
\begin{equation}
\xi_{\parallel}\sim\frac{e^{\Sigma\theta L}}{4\Sigma\theta^2}
\label{+-xi}
\end{equation}
From this we can now extract the desired result for the point 
tension in the interfacial model using the ALU identification (\ref{expo}):
\begin{equation}
\tau=-\ln\theta+A
\label{cwfirst}
\end{equation}
where $A=-\ln 2\sqrt{\Sigma \xi_0}$ may be regarded as an unimportant
non-singular contribution, the value of which depends on the choice of reference
length-scale $\xi_0$. In the interfacial model (\ref{effham}) with short-ranged forces 
the only possible choice of length-scale intrinsic to the interface is the 
inverse surface stiffness so that $\xi_0=1/\Sigma$
which is directly proportional to the bulk correlation length (and recall we
have set $k_B T\equiv 1$). With this
choice of reference length-scale our expression for the point
tension in pure systems is simply
\begin{equation}
\tau(\theta)=-\ln\theta-\ln2
\label{cwpoint}
\end{equation}
In terms of the parameters $\Sigma$ and $\lambda$ this is equivalent to
\begin{equation}
\tau=-\ln\frac{\lambda}{\Sigma}-\ln2
\label{ptlam}
\end{equation}
which will be useful when we consider random-bond systems.

Having derived this result using the ALU identification observe that the 
logarithmic divergence of the point tension for pure systems is also
consistent with the behaviour of the PDF $P_\pi(l;\theta)$ for wetting at a
single wall. To see this
recall that, similar to the correlation length $\xi_{\parallel}$ (for 
the $\bf{+-}$ strip geometry), the scaling form of the PDF $P_\pi(l;\theta)$ is 
also determined by the surface free-energy and point tension. Ignoring the 
normalisation 
constraint for the moment, notice that 
$P_\pi(l;\theta)$ may be identified as 
\begin{equation} 
P_\pi(l;\theta)\propto e^{-f^{\times}(l)}
\label{efx}
\end{equation}
where $f^{\times}(l)$ denotes the excess free-energy cost of an interfacial
configuration constrained to be at height $l$ at some arbitrary position
along the wall (which we can take to be the origin). For asymptotically large
distances $l \gg l_\pi$ typical interfacial configurations determining 
$P_\pi(l;\theta)$ will have a triangular shape with incident angle $\theta$. The free-energy $f^{\times}(l)$ can therefore be estimated as
\begin{equation}
f^{\times}(l)=2\Sigma\theta l+2\tau
\label{fxst}
\end{equation}
showing contributions from two point tensions and the surface free-energies.
Notice this latter term is precisely twice the value of the analogous 
contribution to the ALU correlation length 
(\ref{expo}) so we immediately recover the expression for the PDF 
(\ref{PDFThermal}). To extract the point tension from $P_\pi(l;\theta)$ we have
to bear in mind that unlike the correlation length identification (\ref{expo}) 
the PDF satisfies the addition constraint of normalisation. Moreover in 2D one
is not capable of distinguishing the normalisation constant from the point
tension term $e^{-2\tau}$ since the latter term is simply another constant
(independent of $l$). Turning this around we observe that the normalisation constant
$N(\theta)$ appearing in the asymptotic scaling form
\begin{equation}
P_\pi(l;\theta)=N(\theta)e^{-f^{\times}(l)}
\label{norm}
\end{equation}
must be related to the (exponential) of the point tension. The ALU
identification of $\tau$ through the correlation length $\xi_{\parallel}$ is
in fact in precise accord with the behaviour of the PDF provided we identify
\begin{equation}
P_\pi(l;\theta)=e^{-2\Sigma\theta l-\tau}/\xi_0
\label{enorm}
\end{equation}
or more simply
\begin{equation}
\tau=-\ln\left(N\xi_0\right)
\label{tauN}
\end{equation}
where again $\xi_0$ is an appropriate intrinsic length-scale introduced for
dimensional reasons and which plays no role in determining the divergence of
$\tau$. Using the choice $\xi_0=1/\Sigma$ appropriate to the interfacial 
model we recover the ALU identification (\ref{cwpoint}).

\subsection{The point tension for critical wetting with random bonds}

We now turn to the evaluation of the point tension for 2D critical wetting
with random bond disorder. Together with the interfacial model result for pure 
systems this will be crucial in our discussion of the
scaling connections between filling and the SFL regime of critical wetting. With
random bond disorder the interfacial model for 2D wetting is written
\cite{Forgacs,Kardar}
\begin{equation}
H[l] = \int dx \left\{ \frac{\Sigma}{2} \left(\frac {dl}{dx} \right)^2 + hl+W(l)
+V_r(x,l) \right\}
\label{RBCW}
\end{equation}
where the Gaussian random variable $V_r(x,l)$ has statistical properties
\begin{equation}
\overline{V_r(x,l)}=0
\label{V}
\end {equation}
\begin{equation}
\overline{V_r(x,l)V_r(x',l')}-\overline{V_r(x,l)} \; \overline{V_r(x',l')}=\Delta\delta\left(x-x'\right)\delta\left(l-l'\right)
\label{VV}
\end {equation}
where the overbar denotes an average over the quenched disorder with
strength $\Delta$. It is
convenient to introduce the length-scale $\kappa=\Delta\Sigma/2$ as a 
measure of the bulk disorder which vanishes for the pure (thermal) system. As
first shown by Kardar \cite {Kardar} the model can be studied using the replica trick
identification 
\begin{equation}
\overline{\ln Z_\pi}=\lim_{n\to 0} \frac{\overline{Z_\pi^n}-1}{n}
\label{rep}
\end{equation}
where $Z_\pi^n$ may be interpreted as the partition function for $n$ 
non-interacting
interfaces in an environment with bulk random bonds. Some details of this
calculation are repeated below together with the results necessary for the
calculation of the point tension and later the wedge interfacial height distribution function and
free-energy. Performing the disorder
average introduces interactions described by the many-body Hamiltonian (ignoring
$l$ independent terms)
\begin{equation}
H[\{l_i\}]= \int dx \Big\{\sum_{i=1}^{n} \left( \frac{\Sigma}{2} \left(\frac
{dl_i}{dx} \right)^2 +hl+
W\left(l_i\right)\right) -\Delta\sum_{i<j}^{n}\delta\left(l_i-l_j\right)\Big\}
\label{many}
\end{equation}
so that the interacting n-body partition function for interfaces of length 
$X$ with boundary values ${l_i}$ at $x=0$ and ${l'_i}$ at $x=X$ has the 
spectral expansion
\begin{equation}
\overline{Z_\pi({l_i},{l'_i};X)}=\sum_{m=0}^{\infty}\psi_m^{(n)*}(\{l'_i\})
\psi_m^{(n)}(\{l_i\})e^{-E_m X}
\label{nbod}
\end{equation}
where $\psi_m^{(n)}$ is the $m$-th state wave-function with eigenvalue $E_m$
for $n$-interacting interfaces satisfying the Schr\"odinger equation
$\widehat{H}^{(n)}\psi_m^{(n)}=E_m\psi_m^{(n)}$. Again, ignoring constant terms, the
Hamiltonian operator is 
\begin{equation}
\widehat{H}^{(n)}=-\sum_{i=1}^{n} \left( \frac{1}{2\Sigma}\frac{\partial^2}{\partial
l_i^2}+hl+W(l) \right) -\Delta\sum_{i<j}^{n}\delta(l_i-l_j)
 \label{HamN}
\end{equation}
For systems with strictly short-ranged forces, characteristic of the SFL regime, it is convenient to adopt the 
natural generalisation of the boundary condition (\ref{bound1}) which reads
\cite{Forgacs}
\begin{equation}
\lim_{l_i\to 0}\frac{\partial
\psi_m^{(n)}(\{l_j\})}{\partial l_i}=-\lambda\psi_m^{(n)}(\{l_j\}) \Big|_{l_i=0}
\label{boundRB}
\end{equation}
or any $l_j$. Note that the length-scale $\lambda$ is characteristic of the pure
wall-fluid interface and remains finite at the wetting transition in the
presence of random-bonds. The groundstate solution to the eigenvalue problem is given
by the Bethe-$\it{ansatz}$ wavefunction \cite {Forgacs,Kardar}
\begin{equation}
\psi_0^{(n)}(\{l_i\})=C_n(\lambda,\kappa)e^{-\lambda\sum l_i+\kappa\sum_{i<j}\vert
l_i-l_j\vert}
\label{bethe}
\end{equation}
with normalisation constant
\begin{equation}
C_n(\lambda,\kappa)=(2\kappa)^{n/2}\Bigg(\frac{\Gamma(\lambda/\kappa+2n-1)}
{\Gamma(\lambda/\kappa+n-1)}\Bigg)^{1/2}
\label{NormBethe}
\end{equation}
where $\Gamma(x)$ is the usual gamma function. From the wave-function one
can easily obtain the groundstate energy, and by considering the limit of
$E_0^{(n)}/n$ as $n\to 0$ identify 
\begin{equation} 
f_{\mathrm{sing}}=-\frac{(\lambda-\kappa)^2}{2\Sigma}
\label{RBF}
\end{equation}
as the singular contribution to the free-energy. Thus the contact angle is simply
\begin{equation}
\theta=\frac{(\lambda-\kappa)}{\Sigma}
\label{RBcontact}
\end{equation}
and note that these expressions identically reproduce the results for the pure
system when $\kappa\to 0$. From the above it is clear that the disorder lowers
the wetting transition temperature which now occurs at $\lambda=\kappa$. We shall also need the expression for the
mean interface height 
\begin{equation}
l_\pi=\frac{1}{2\kappa}\psi'\Big(\frac{\lambda}{\kappa}-1\Big)
\label{lsum}
\end{equation}
which involves the derivative of the psi or digamma function defined by 
\begin{equation}
\psi(x)=\frac{d\ln\Gamma(x)}{dx}
\label{phi}
\end{equation}
Again in the limit $\kappa\to 0$ this reproduces the
 appropriate result $l_\pi=1/(2\lambda)$ for pure systems. For finite $\kappa$ however the asymptotic divergence of $l_\pi$ as
$\lambda\to\kappa$ is different to the pure system and
\begin{equation}
l_\pi\sim\frac{\kappa}{2(\lambda-\kappa)^2}
\label{RBl}
\end{equation}
From the results for the singular contribution to the free energy and divergence of
the interfacial height we have $\alpha_s=0$ and $\beta_s=2$ in agreement with
the general expectations for the SFL regime with $\zeta=2/3$. The 
PDF $P_\pi(l;\theta)$
describing the fluctuations of the interfacial height in the asymptotic scaling
regime is given by (\ref{PDFRB}).

To evaluate the point tension for random-bonds we use the properties of the
probability distribution function taking care to extract the relevant
quantities at finite $n$ before continuing to $n=0$. The $n-$point PDF is the
square of the groundstate wavefunction which may be written as the ordered
product
\begin{equation}
P^{(n)}(\{l_i\})=C_n^2(\lambda,\kappa)\prod_{j=1}^{n}e^{-2(\lambda+(n+1-2j)\kappa)l_j}
\label{ordered}
\end{equation}
with $l_1<l_2<..<l_n$. By analogy with the interpretation of the PDF for pure 
systems the coefficient of each
 $l_j$ term  appearing in the exponential may be viewed as the surface
free-energy cost of constraining the height of the $j^{th}$ interface whilst
 the normalisation constant contains the required information about the point
tension. Using the appropriate replica trick identification we generalise the result 
(\ref{tauN}) for the point tension in the pure system to 
\begin{equation}
\tau=-\lim_{n\to 0}\frac{1}{n}\Big(C_n^2(\lambda,\kappa)(\xi_0^{RB})^n-1\Big)
\label{DefTauR}
\end{equation}
where, in an obvious notation $\xi_0^{RB}$ is a suitable choice of reference length-scale for the
random-bond system which plays the same, trivial dimensional role as the
length-scale $\xi_0$ for systems with purely thermal disorder. We emphasise that 
the choice of $\xi_0^{RB}$ does not 
influence the asymptotic divergence of the point tensions as $T\to T_{\mathrm{wet}}$ and
only contributes towards the non-singular, background term analogous to the
constant $A$ appearing in (\ref{cwpoint}). Thus we find
\begin{equation}
\tau=-\psi\Bigg(\frac{\lambda}{\kappa}-1\Bigg)-\ln2\kappa\xi_0^{RB}
\label{Tphi}
\end{equation}
which again introduces the digamma function. We now choose the
value of $\xi_0^{RB}$ so that upon taking the limit $\kappa\to 0$ we recover
the correct background term for the pure system (\ref{ptlam}). As $\kappa\to 0$
the argument of the digamma function diverges and we can use the asymptotic
large $x$ expansion
\begin{equation}
\psi(x)\sim\ln x-\frac{1}{2x}+....
\label{asphi}
\end{equation}
Note that the necessary logarithmic singularity for the point tension in the
pure system emerges naturally from the
properties of the digamma function. The appropriate choice of reference
length-scale is therefore $\xi_0^{RB}=(\Sigma)^{-1}\equiv\xi_0$ and is
$\it{identical}$ to that chosen in our
earlier discussion of purely thermal disorder. We regard this as a rather
pleasing feature of the present replica trick definition of $\tau$ using
the PDF. In terms of
the length-scales $\lambda$, $\kappa$ and $\Sigma$ our expression for the point tension
with random-bond disorder is therefore
\begin{equation}
\tau=-\psi\Big(\frac{\lambda}{\kappa}-1\Big)+\ln\Big(\frac{\Sigma}{2\kappa}\Big)
\label{TauR}
\end{equation}
Alternatively for fixed $\Sigma$ and $\kappa$ we can eliminate $\lambda$ and
rewrite this in terms of the contact angle $\theta$ recalling that
$\theta\propto\lambda-\kappa\propto t'$. This is the form
that is most convenient for the discussing the connection with 2D filling. Our
final result is
\begin{equation}
\tau(\theta)=-\psi\Big(\frac{\theta\Sigma}{\kappa}\Big)+\ln\Big(\frac{\Sigma}{2\kappa}\Big)
\label{Taufinal}
\end{equation}
which should be compared with (\ref{cwpoint}) for the pure system. Equations
(\ref{TauR}) and (\ref{Taufinal}) are the main new results of this section and will
play an important role in our discussion of 2D wedge filling with
random-bond disorder. 

We are now in a position to test the validity of the
Indekeu-Robledo critical exponent relation for the line/point tension. The
singularities of the point tension occurring as $\theta\to 0$ are contained
within the digamma function which diverges as $\psi\sim-1/x$ as $x\to 0$. 
Thus we can identify the singular contribution to the point tension
\begin{equation}
\tau_{\mathrm{sing}}\sim\frac{\kappa}{\Sigma\theta}
\label{RTsing}
\end{equation}
implying $\alpha_l=3$ which is in precise agreement with the conjectured exponent
relation. Note also that as with the pure system the point tension diverges to
$+\infty$ as $T\to T_{\mathrm{wet}}$ although the quantitative divergence is much stronger.

\section{Two dimensional filling in pure and impure systems I: Scaling theory}

Our presentation of fluctuation effects occurring at 2D filling transitions
parallels our earlier treatment of critical wetting. In turn we will consider
 (A) the
definitions of critical exponents and the derivation of exponent relations, (B)  
a discussion of fluctuation regimes from heuristic scaling arguments and (C)
the scaling and SDE of the density profile and PDF. From these preliminary
considerations  will emerge a possible fluctuation-induced connection with the
 SFL regime of critical wetting which will be precised later using covariance 
relations.

\subsection{Critical exponents and exponent relations}

A 2D wedge is a `V' shaped substrate formed 
from the junction of two planar (identical) walls that meet at
the origin (say) with angles $+\alpha$ and $-\alpha$ measured w.r.t to the 
$z=0$ line. Thus the height of the wall above
the line is described by a wall-function $z_w(x)=\cot\alpha\vert x\vert$,
where the $x$-axis runs across the wedge. The wedge is considered to be in contact with a 
bulk vapour phase at temperature $T$ and chemical potential $\mu$ and is
supposed to preferentially adsorb the liquid phase along the surface of the
walls and, in particular, the wedge bottom. Thus the equilibrium density profile
$\rho(z,x)$ is liquid-like in the filled region and will show packing
effects very close to the wall although these will not be our concern here. Very general macroscopic,
thermodynamic arguments \cite{Finn,Pomeau,Hauge} indicate that at bulk coexistence $\mu=\mu_{\mathrm{sat}}$ the 
wedge is completely filled by liquid provided the planar contact angle
satisfies $\theta<\alpha$. Thus for the most common case where the contact
angle decreases with increasing temperature, the transition from partial to
complete filling occurs at a filling transition temperature $T_{\mathrm{fill}}$ satisfying
\begin{equation}
\theta(T_{\mathrm{fill}})=\alpha
\label{Tf}
\end{equation}
This implies that complete filling precedes complete wetting and also that the
filling temperature $T_{\mathrm{fill}}$ can be lowered simply by increasing
the angle of the wedge. On approaching the filling phase boundary at
$(T_{\mathrm{fill}},\mu_{\mathrm{sat}}(T_{\mathrm{fill}}))$ the mean height of the interface $l_w$,
as measured from the wedge bottom, diverges. The divergence is discontinuous and continuous for
first-order and second-order (critical) filling respectively. Whilst in 3D both
types of transition are possible, in 2D filling transitions will almost 
always be continuous. At two-phase coexistence the equilibrium height profile 
$l_{eq}(x)$ measured from the $z=0$ line, with $l_w\equiv l_{eq}(0)$ is essentially flat in the filled region of the wedge 
owing to the absence of any macroscopic curvature as dictated by the Laplace
equation. The lateral extent of the filled region is therefore controlled by a
correlation length $\xi_x\approx 2l_w\cot \alpha$ which is trivially related to
the interfacial height. Critical effects at 2D filling may be viewed as
arising from breather-mode-like fluctuations in the interface height which roll
the points of contact up and down the sides of the wedge thus changing the
height and volume of the filled region. A similar picture
holds for 3D conic filling but is modified in a 3D wedge owing to
long-wavelength fluctuations along the system. Similar to wetting the filling transition has two 
relevant scaling fields
which we can write $t=(T_{\mathrm{fill}}-T)/T_{\mathrm{fill}}$ and $h=(\rho_l-\rho_v)(\mu_{\mathrm{sat}}-\mu)$ respectively. However for
filling one has the additional possibility of using the wedge angle to control
the deviation from the phase boundary. Thus for fixed $T$ close to $T_{\mathrm{fill}}$ the
combination $\theta-\alpha \propto t$ is a linear measure of the temperature-like
scaling variable. At bulk coexistence $\mu=\mu_{\mathrm{sat}}$ the divergence of the
mid-point interfacial-height and roughness are characterised by critical
exponents
\begin{equation}
l_w\sim t^{-\beta_w},\xi_{\perp}\sim t^{-\nu_{\perp}}
\label{lf}
\end{equation}
and we anticipate that in a fluctuation-dominated regime $\beta_w=\nu_{\perp}$ so
that $l_w\sim \xi_{\perp}$. Along the filling critical isotherm 
$T=T_{\mathrm{fill}}$, $h\to 0$, the midpoint
height $l_w$ (and $\xi_{\perp}$) also diverges and we introduce the critical 
exponent 
\begin{equation}
l_w\sim h^{-\psi_w}
\label{lhww}
\end{equation}
to characterise this. Two other critical exponents are defined from the
singularities of the wedge free-energy $f_w(\theta,\alpha,h)$ which, for later
purposes, we have written as a function of the variables which highlight the
covariance with wetting. Also for $h=0$ we define $f_w(\theta,\alpha)\equiv 
f_w(\theta,\alpha,0)$. At a thermodynamic level the wedge
free-energy is defined by subtracting from the total grand potential $\Omega$
the bulk free-energy and the contribution from two (infinite) planar walls:
\begin{equation} 
f_w(\theta,\alpha,h)=\Omega+pV-\sigma^{(\pi)}_{wv} A
\label{fw}
\end{equation}
where $\sigma^{(\pi)}_{wv}$ is the wall-vapour tension for the planar 
($\alpha=0$) system
and $A$ is the $\it{total}$ surface area exposed to fluid. By construction the wedge
free-energy vanishes in the planar limit $\alpha=0$. On the other hand we
 expect that $f_w(\alpha,\alpha)$ is unbounded due to the
adsorption of a macroscopic amount of liquid. Near the filling transition
we anticipate that $f_w$ contains a singular contribution that the shows 
scaling behaviour depending on the variables $h$ and $t\propto (\theta-\alpha)$
only. We write 
\begin{equation}
f^{\mathrm{\mathrm{sing}}}_w \sim t^{2-\alpha_w} W_w(ht^{-\Delta_w})
\label{wsing}
\end{equation}
which introduces the wedge specific heat exponent $\alpha_w$, gap
exponent $\Delta_w$ and free-energy scaling function $W_w(x)$. Partial derivatives of the wedge free-energy are related
to thermodynamic observables, similar to the Gibbs adsorption equation for
planar systems. Firstly, in the free-energy the bulk ordering-field $h$ is conjugate to the total 2D volume of
adsorbed fluid so that 
\begin{equation}
\frac{\partial f^{\mathrm{sing}}_w}{\partial h}\propto l_w^2
\label{Maxh}
\end{equation}
Secondly variation of the wedge angle $\alpha$ linearly changes the height and 
lateral extent of the filled region, implying
\begin{equation}
\frac{\partial f^{\mathrm{sing}}_w}{\partial t}\propto l_w
\label{Maxt}
\end{equation} 
In this way we obtain the exponent relations
\begin{equation}
\Delta_w=2-\alpha_w+2\beta_w
\label{delf}
\end{equation}
and 
\begin{equation}
1-\alpha_w=-\beta_w
\label{bef}
\end{equation}
showing there is only one free critical exponent for 2D filling. 

\subsection{Fluctuation regimes for 2D filling}

  The classification of fluctuation regimes and also the values of the critical
exponents for 2D filling follow from a rather simple heuristic scaling theory
somewhat analogous to the Lipowsky-Fisher treatment of critical wetting
considered earlier. To begin we
consider mean-field theory which ignores the fluctuation effects arising from
thermal excitations or quenched impurities. As first shown by Rejmer \etal \cite{MFT}, interfacial models give a very elegant description of filling 
phenomena at mean-field level. For open wedges corresponding to
 small $\alpha$ (for which $\tan \alpha\approx \alpha$) the equilibrium
mean-field profile $l_{eq}(x)$ may be found from minimisation of the effective
interfacial free-energy \cite{MFT}
\begin{equation} 
F_w[l] = \int dx \left\{ \frac{\Sigma}{2}\left(\frac {dl}{dx}\right)^2 
+h(l-\alpha \vert x \vert)+W(l-\alpha \vert x \vert)\right\}
\label{Feffwedge}
\end{equation}
which can be justified from analysis of a more general drumhead-like
model valid for larger $\alpha$. As mentioned earlier, $l(x)$ denotes the interfacial
height relative to the $z=0$ line, whilst $W(l)$ is the binding potential
appropriate to the $\it{planar}$ system. We emphasise that the small $\alpha$ 
approximation is not expected to 
introduce any peculiarities and the critical behaviour predicted by the model 
(at mean-field level and beyond) is believed to be valid for arbitrary wedge 
angles. The free-energy functional is minimised subject to the appropriate boundary conditions
that the equilibrium profile $l_{eq}(x)\to \alpha\vert x \vert+ \; l_{\pi}$ as
$\vert x \vert \to \infty$. The resulting Euler-Lagrange equation can be
integrated once to give an explicit equation for the mid-point height
(restricting our attention to $h=0$) \cite{MFT,ourJphysCM,ourPRL2},
\begin{equation}
\frac{\Sigma(\alpha^2-\theta^2)}{2}=W(l_w)
\label{Euler}
\end{equation}
which can solved trivially. Thus for binding potentials of the form (\ref{BPot}) the dependence on $q$ is
unimportant and the mean-field critical behaviour is determined solely by the
leading order index $p$. As $\theta\to\alpha$ we may expand the above equation
\begin{equation}
\theta=\alpha+\frac{a}{\alpha\Sigma}l_w^{-p}+\dots
\label{EULasymp}
\end{equation}
recalling that $a$ remains $\it{finite}$ at the filling transitions since
$T_{\mathrm{fill}}<T_{\mathrm{wet}}$. From this it follows immediately that the mean-field value 
of the height critical exponent at filling is $\beta_w=1/p$ \cite{ourPRL2}.
The absence of any $q$ dependence is a first indication that critical filling
 may be less sensitive to
the nature of the intermolecular forces compared to critical wetting.

It is possible to extend this simple mean-field approach to include thermal and
disorder-induced fluctuation
effects in a heuristic way. The last equation tells us how the
difference or shift between the contact angle and wedge tilt angle depends on the 
$\it{direct}$ influence of the intermolecular forces. In the presence of
fluctuation effects arising from either thermal or random-bond disorder it is
natural to suppose that (\ref{EULasymp}) generalises to
\begin{equation}
\theta=\alpha+\Delta\alpha_{p}(l_w)+\Delta\alpha_{fl}(l_w)
\label{anglesum}
\end{equation}
where $\Delta\alpha_{p}(l)\sim l^{-p}$ is the direct angle shift and 
$\Delta\alpha_{fl}(l)$ is the shift arising due 
to fluctuation effects which we anticipate takes the form of a ratio of
length-scales. Now by construction $\Delta\alpha_{fl}(l)$ is
$\it{not}$ the ratio $l_w/\xi_{x}$  since this is, essentially, the first
term in (\ref{anglesum}) and is purely geometrical. Instead we write 
$\Delta\alpha_{fl}(l)\propto\xi_{\perp}/\xi_{fl}$ where $\xi_{fl}$ is 
an appropriate fluctuation-related length-scale which must be much
larger than the mid-point height. If we also make the reasonable assumption
that this length-scale is controlled by the wandering exponent $\zeta$ then the 
 simplest 
possible choice given these constraints is $\xi_{fl}\sim l_w^{1/\zeta}$ similar to the relation 
$\xi_{\parallel}\sim l^{1/\zeta}$ appropriate to a fluctuation-dominated (WFL or SFL
regime) wetting
transition. Consequently if fluctuations dominate we anticipate
\begin{equation}
\Delta\alpha_{fl}(l)\sim l^{1-1/\zeta}
\label{delfl}
\end{equation}
similar to the length-scale ratio 
$\xi_{\perp}/\xi_{\parallel}$ (\ref{precursor}) appearing in the 
Lipowsky-Fisher analysis.  
  
This simple, heuristic modification of the mean-field analysis is
particularly powerful because for filling the phase boundary always remains
$\theta=\alpha$ and is not modified by fluctuations. It follows that the
critical behaviour should fall into two possible classes:
\begin{itemize}
\item{{\bf Filling mean field (FMF) regime} - if $p<1/\zeta-1$ fluctuation effects are
negligible, $l_w \gg \xi_{\perp}$ and the critical exponent $\beta_w=1/p$ is
 unchanged from its mean-field value.}
\item{{\bf Filling fluctuation (FFL) regime} - if $p>1/\zeta-1$ there are large-scale
fluctuations, $l_f\sim \xi_{\perp}$, and the critical exponents are universal
and determined by the wandering exponent. For the divergence of the filling 
height we predict
\begin{equation}
\beta_w=\frac{\zeta}{1-\zeta}
\label{bf}
\end{equation}
with the values of the other critical exponents following from the 
relations (\ref{delf}) and (\ref{bef}).}
\end{itemize}
At this point, a number of remarks are in order:
\begin{itemize}
\item[(I)] These predictions are in perfect agreement with exact
results known from transfer-matrix and replica trick studies of interfacial
models which find $\beta_w=1$ and $\beta_w=2$ for pure ($\zeta=1/2$) and impure
($\zeta=2/3$) systems with short-ranged forces \cite{ourPRL1,ourJphysCM}. They
are also consistent with studies of filling (corner wetting) in
square lattice Ising models \cite{ourPRL3,Dux,Adam}.
Moreover for pure systems it is possible to completely classify the critical
behaviour using the interfacial model \cite{ourPRL1} and show that the criticality falls into the above 
two regimes with a marginal value $p=1$ corresponding to the FFL/MF 
borderline. The critical exponent remains $\beta_w=1$ for this marginal case.\\
\item[(II)] The existence of two fluctuation regimes for filling clearly contrasts with the
phenomenology of critical wetting for which there
 are three. Also note that the borderline between the FFL and FMF regime occurs
when $p=1/\zeta-1$ which is different to the SFL/WFL and WFL/MF borderlines for
critical wetting which happen when $p=2(1/\zeta-1)$ and $q=2(1/\zeta-1)$
respectively. The regime in which there is universal critical behaviour is 
broader for filling than for wetting.\\
\item[(III)] The value of the critical exponent $\beta_w=1/p$ in the FMF regime is 
different to the value $\beta_s=1/(q-p)$ in the MF regime of critical wetting.
Therefore when the intermolecular forces are sufficiently long-ranged to induce
induce mean-field-like criticality, there is no apparent connection between filling 
and wetting. However for sufficiently short-ranged forces the predicted value
for the critical exponent $\beta_w$ (\ref{bf}), belonging to the FFL
regime, is the same as the random-walk result for the critical exponent $\beta_s$ for 
the critical wetting SFL regime (\ref{SFL1}). This is a first hint that there 
may be some fluctuation-induced connection between the two transitions.
\end{itemize}

\subsection{Scaling of the PDF and short-distance expansion}
 
In the FFL regime we anticipate that, within the filled region of the wedge, 
the density profile $\rho(z,x)$ exhibits universal scaling behaviour related 
to the scaling of the interfacial height PDF. We will focus on the behaviour of the density profile and
distribution function occurring at the centre of the wedge ($x=0$) and define
$\rho_w(z)\equiv\rho(z,0)$. Thus, for $z\to\infty$, $t\to 0$, $h\to 0$ with
$zt^{\beta_w}$ and $ht^{-\Delta_w}$ arbitrary we expect 
\begin{equation}
\rho_w(z)=\rho_l-(\rho_l-\rho_v) \Xi_w(zt^{\beta_w},ht^{-\Delta_w})
\label{rhof}
\end{equation}
where the scaling function satisfies $\Xi_w(\infty,y)=1$ and $\Xi_w(0,y)=0$ for 
any $y$. Notice that unlike the case of critical wetting, where one has to distinguish
between scaling behaviour in the SFL and WFL regimes, the scaling function for 
fluctuation-dominated filling is unique. Associated with the scaling of the
profile is a SDE describing the algebraic behaviour close to the
wall compared to the filling height. At $h=0$ we write, analogous to (\ref{gamma})
\begin{equation}
\rho_w(z)-\rho_l \approx (\rho_v-\rho_l)(zt'^{\beta_w})^{\gamma_w}
\label{gammaf}
\end{equation}
which introduces our final critical exponent ${\gamma_w}$ for filling and which is only
defined for the FFL regime. Similar to SDE
exponents for SFL and WFL regime wetting, the value of the critical 
exponent $\gamma_w$ is not independent and can be related to the other exponents
defined for filling. To see this, consider that the value of the wall-fluid
intermolecular potential contains an additional short-ranged contribution of
strength $h_0$ localised to the bottom of the wedge. In a magnetic (Ising)
language this is would correspond to an incremental point field at the wedge
apex and serves only to introduce a new non-singular length-scale proportional
to the value of the field. This is useful because differentiation of the 
wedge free-energy w.r.t. $h_0$ yields the value of
the density at or near the wedge bottom. Now
the field $h_0$ is irrelevant, in the renormalisation group sense, and can be
included in the scaling hypothesis for the free-energy (\ref{wsing}) by allowing
for an additional scaling variable $h_0 t^{\beta_w}$ which is simply the ratio
of relevant length-scales. Differentiation of the singular contribution to the
wedge free-energy therefore implies that the singular contribution to
the density at the wedge bottom is simply $\rho_w(0)\sim t$ where we have used
the exponent relation (\ref{bef}). However from the SDE we can also identify 
$\rho_w(0)\sim t^{\beta_w\gamma_w}$ implying that
\begin{equation}
\gamma_w=1/\beta_w
\label{SDEW}
\end{equation}
which will later prove to be an extremely useful exponent relation.

The scaling of the profile at filling follows from the scaling of the PDF for
the mid-point interfacial height, written $P_w(l;\theta,\alpha,h)$, similar to
(\ref{density}). At bulk coexistence, $h=0$, we simply write
\begin{equation}
P_w(l;\theta,\alpha,0)\equiv P_w(l;\theta,\alpha)
\label{pf}
\end{equation}
In the FFL regime we expect that $P_w(l;\theta,\alpha)$ is characterised by a
universal scaling function $\Lambda_w(x)$ such that
\begin{equation}
P_w(l;\theta,\alpha)=\tilde{a}(\theta-\alpha)^{\beta_w}\Lambda_w(\tilde{a}l(\theta-\alpha)^{\beta_w})
\label{atlast}
\end{equation}
where $\Lambda_w(x)$ is a universal function and the inverse length-scale
$\tilde{a}$ is chosen, as with $\Lambda_{\pi}(x)$, so that the argument is
simply $l/l_w$. Clearly the PDF has the SDE $\Lambda_w (x) \sim x^{\gamma_w-1}$. The 
relationship between the universal scaling function $\Lambda_w(x)$ for filling
and the corresponding function $\Lambda_{\pi}^{SFL}(x)$ for 2D wetting will be 
central to our study.

\section{Interfacial models of 2D filling (II): Exact results and covariance}

\subsection{Transfer matrix results} 

 We begin with the transfer matrix theory of filling in pure systems
\cite{ourPRL1} based on
the interfacial Hamiltonian
\begin{equation} 
H_w[l] = \int dx \left\{ \frac{\Sigma}{2}\left(\frac {dl}{dx}\right)^2 +h(l-\alpha \vert x \vert)+
W(l-\alpha \vert x \vert) \right\}
\label{effwedge}
\end{equation}
valid for open wedges. It is easiest to assume that the
horizontal range is $[-X/2,X/2]$  with periodic boundary conditions at the
end-points. Note that
the model trivially recovers the interfacial Hamiltonian for planar wetting $H[l]$ (\ref{effham})
when $\alpha=0$. Again we emphasise that the
assumption of small $\alpha$ is not believed to be in any way important as
regards the critical behaviour occurring near the filling transition and
predictions based on the above interfacial model are supported by Ising model 
studies of filling at right-angle corners for different lattice types
\cite{ourPRL3}. To
obtain the partition function corresponding to the fluctuation sum over
Boltzmann weights it is convenient to make the change of variable
 $\tilde l\equiv l-\alpha \vert x \vert$ in which case we can re-write the
Hamiltonian as
\begin{equation}
H_w=2\Sigma \alpha \left( \tilde{l}_e-\tilde{l} (0) \right) +H[\tilde l]
\label{tilde}
\end{equation}
where $\tilde l_e \equiv l(X/2)$ denotes the end-point interfacial height
(relative to the wall) and $\tilde l(0)\equiv l(0)$ is the mid-pont height
above the bottom of the wedge. Thus the angle $\alpha$ enters the partition
function only through a local
exponential boost factor associated with the mid-point height (and end-points).
The ensemble average $\langle l(0) \rangle$ defines the equilibrium mid-point 
height $l_w$ and from (\ref{tilde}) it
is immediately apparent that
\begin{equation}
l_w=-\frac{1}{2\Sigma}\frac{\partial f_w}{\partial \alpha}
\label{exact}
\end{equation}
which is a precise version of (\ref{Maxt}). The same relation is also valid in the presence of
random-bond disorder and will prove useful later. The model can be analysed
using continuum transfer-matrix methods which yield very general
expressions for the wedge free-energy and interfacial height PDF, valid for
general choices of binding potential. In the thermodynamic limit $X \to \infty$
and in terms of the inner product defined 
in (\ref{matrix}) the wedge free-energy follows as
\begin{equation}
f_w(\alpha,\theta,h)=-\ln \langle 0 \vert e^{2\Sigma\alpha l} \vert 0\rangle
\label{00}
\end{equation}
The probability of finding the interface at height $\tilde{l}$ from the wall at
position $x$ along it is given by
\begin{equation}
P_w(\tilde{l},x)=\sum_n\frac{\langle n \vert e^{2\Sigma\alpha l}  \vert
0\rangle \psi_n^*(\tilde{l})\psi_0(\tilde{l})e^{(E_0-E_n)\vert x \vert}}{\langle 0 \vert
e^{2\Sigma\alpha l}  \vert 0\rangle}
\label{pxl}
\end{equation}
requiring knowledge of the full transfer matrix spectrum of the planar system.
At the mid-point ($x=0$) however, for which $\tilde{l}=l(0)$, symmetry
considerations simplify the expression considerably and
\begin{equation}
P_w(\tilde{l},0)\equiv P_w(l)= \frac{ \vert \psi_0(l) \vert ^2 e^{2\Sigma\alpha l}}{\langle 0 \vert
e^{2\Sigma\alpha l}  \vert 0\rangle}
\label{p}
\end{equation}
 which only depends on the groundstate properties of the planar problem. This
is indicative that the mid-point PDF will play a special role in the theory of
wedge filling. Using these relations it is easy to establish that the filling transition is
located at $\theta=\alpha$ (and $h=0$) in precise accord with the thermodynamic
prediction. Moreover the critical behaviour falls into two categories in
agreement with the heuristic treatment of the previous section. For binding
potentials with $p>1$ the asymptotic criticality is mean-field-like with
$\beta_w=1/p$ and $\nu_{\perp}=(1+p)/2p$ so that $l_w\gg\xi_{\perp}$. In the FFL regime corresponding to $p>1$ the
behaviour in the asymptotic scaling regime is universal and the same as that
found for systems with purely short-ranged forces using the
boundary-conditions (\ref{bound1}). At $h=0$ the scaling expressions pertinent
to this critical regime are 
\begin{equation}
 l_w=\frac{1}{2\Sigma(\theta-\alpha)}
\label{1}
\end{equation}
\begin{equation}
 P_w(l;\theta-\alpha)=2\Sigma(\theta-\alpha) e^{-2\Sigma(\theta-\alpha) l}
\label{2}
\end{equation}
\begin{equation}
f_w=\ln(\theta-\alpha)-\ln\theta
\label{3}
\end{equation}
corresponding to critical exponents $\beta_w=1$ and $2-\alpha_w=0 (\ln)$. 

In the presence of random bond-disorder the generalisation of the
interfacial model (\ref{RBCW}) for filling transitions in open wedges is
\begin{equation}
H_w[l] = \int dx \left\{ \frac{\Sigma}{2} \left(\frac {dl}{dx} \right)^2 +h(l-\alpha \vert x \vert)
+W(l-\alpha \vert x \vert)
+V_r(x,l) \right\}
\label{lasth}
\end{equation}
and for systems with purely short-ranged forces (and at coexistence $h=0$) the 
model can be solved exactly by extending Kardar's replica trick theory
described earlier \cite{ourJphysCM}. The extension is possible because, similar
to (\ref{tilde}) the
replicated Hamiltonian is the same as the corresponding wetting model apart
from a sum over terms $2\Sigma\alpha(\tilde l_i(0)-\tilde l_e)$ which
may be absorbed into the Bethe ansatz (\cite{ourJphysCM}). We omit the details
and only quote the final results for the infinite wedge. The filling transition occurs at
\begin{equation}
 \lambda=\kappa+\Sigma\alpha
\label{Fpb}
\end{equation}
which by virtue of (\ref{RBcontact}) is equivalent to the condition 
$\theta=\alpha$. The mean mid-point height is given exactly by
\begin{equation}
l_w=\frac{1}{2\kappa}\psi'\Big(\frac{(\lambda-\Sigma\alpha)}{\kappa}-1\Big)
\label{lfsum}
\end{equation}
which recovers the pure result (\ref{1}) in the limit $\kappa \to 0$. As
$\theta \to \alpha$ at finite $\kappa$, the interfacial height diverges as 
\begin{equation}
l_w\sim\frac{\kappa}{2(\lambda-\Sigma\alpha-\kappa)^2}
\label{RBlf}
\end{equation}
equivalent to $l_w\sim 1/(\theta-\alpha)^2$ and implying that $\beta_w=2$. The scaling form of the PDF
describing the asymptotic divergence of the $l_w$ is
\begin{equation}
P_w(l;\theta,\alpha)=\frac{\Sigma(\theta-\alpha)}{\pi\sqrt{2l\kappa}}e^{-l(\theta-\alpha)^2\Sigma^2/2\kappa}
\int_{0}^{\infty}ds\frac{\sqrt{s}e^{-s/4}}{s+2l(\theta-\alpha)^2\Sigma^2/\kappa}
\label{FPDFRB}
\end{equation}
Finally the wedge free-energy (at $h=0$) is given exactly by
\begin{equation}
f_w(\theta,\alpha)=\psi \left( \frac{\Sigma(\theta-\alpha)}{\kappa}
\right) - \psi \left(\frac{\theta\Sigma}{\kappa} \right)
\label{ranwedgefree}
\end{equation}
which exactly recovers the pure result (\ref{3}) 
as $\kappa \to 0$. As
$\theta \to \alpha$ the free-energy shows the singular behaviour
\begin{equation}
f_w^{\mathrm{sing}}\sim-\frac{\kappa}{\Sigma(\theta-\alpha)}
\label{RFsing}
\end{equation}
implying $\alpha_w=3$. For both pure and impure systems the wedge free-energy
diverges to $-\infty$ as $\theta\to\alpha$.  

\subsection{Covariance laws for filling and wetting}

The above results for 2D fluctuation-dominated filling in pure and impure
systems point to a remarkable connection with the scaling behaviour occurring 
for the SFL regime of critical wetting. This goes far beyond the identity of the
exponents $\beta_w$ and $\beta_s$ suggested by the heuristic scaling theory. For systems with strictly short-ranged
forces and at bulk coexistence ($h=0$) we have established the following
 covariance relations
\begin{equation}
l_w(\theta,\alpha)=l_\pi(\theta-\alpha)
\label{law0}
\end{equation}
\begin{equation}
P_w(l;\theta,\alpha)=P_{\pi}(l;\theta-\alpha)
\label{law1}
\end{equation}
\begin{equation}
f_w(\theta,\alpha)=\tau(\theta)-\tau(\theta-\alpha)
\label{law2}
\end{equation}
The final relation between the wedge free-energy and the point tension has not
been reported before and is one of the central new results of our paper. 
These `laws' are also valid in the asymptotic critical region, $\theta \to
\alpha$, even in the presence of long-ranged forces provided the filling
transition belongs to the FFL regime. We emphasise that the connection between
filling and the critical wetting SFL regime is all the more remarkable because 
the FFL regime is broader. For
example, recall that with purely thermal disorder the FFL corresponds to binding
potentials with $p>1$ whilst the critical wetting SFL regime corresponds to
$p>2$. Thus for model systems with $1<p<2$ the filling transition precisely
mimics the properties of the SFL regime even though the wetting transition for
the corresponding planar system belongs to the WFL regime. It is in this
sense that the
wedge geometry effectively turns off the influence of the long-ranged forces.
We conjecture that the above covariance relations laws connecting filling and 
wetting are generally true in 2D provided the wandering exponent $1\ge\zeta\ge 1/2$.

\begin{figure}
\begin{center}\resizebox{0.9\textwidth}{!}{%
  \includegraphics{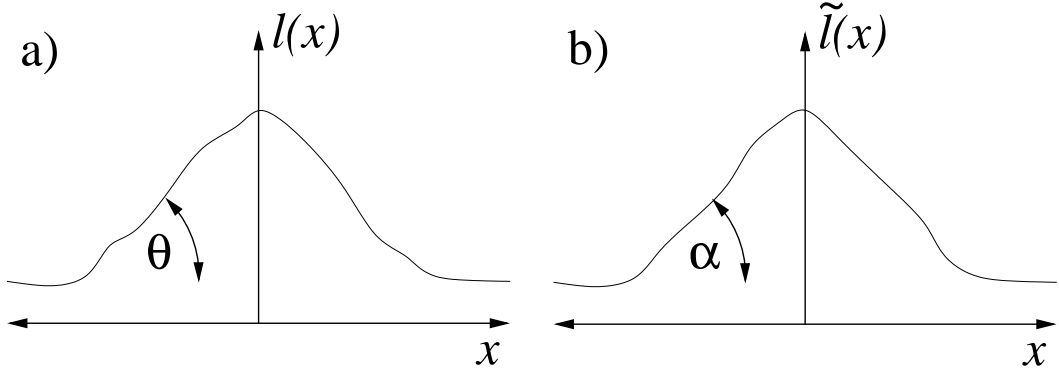}
}
\caption{Interfacial configurations contributing to the one-point
interfacial height probability distribution function in two different
geometries. Figure a) shows the triangular-like configuration an
interface adopts when it is constrained to pass through some arbitrary
point at height $l \gg l_{\pi}$ at bulk two-phase co-existence. Figure
b) shows an interfacial configuration in a two dimensional wedge
geometry represented in terms of the relative height $\tilde{l}$ (see
text). Near a filling transition $\theta \approx \alpha$ and the
contributing profiles to the respective PDF's, $P_{\pi}(l)$ and $P_{w}(l)$,
are essentially the same.}
\label{2rhinos}
\end{center}
\end{figure} 

The fluctuation-induced covariance between filling and wetting has a simple 
geometric interpretation. In \fref{2rhinos} are shown typical
interfacial configurations contributing to the PDF for two different 
geometries. On the LHS is shown the typical triangular 
configuration an interface adopts at a planar wall when it is constrained to 
pass through a point at height $l$ far in excess of the mean interfacial 
height $l_\pi$. On the RHS is the typical configuration for an interface in a 2D wedge geometry
plotted in terms of the relative height 
$\tilde{l}\equiv l(x)-\alpha \vert x \vert$. Close
to the filling transition $\theta\approx\alpha$ and consequently the typical
interfacial fluctuations contributing to the $P_f(l)$ and $P_\pi(l)$ in the 
different geometries are essentially the same. Similar remarks also apply in 3D
for the cone geometry \cite{ourJphysCM2}.

The covariance relations are extremely restrictive and contain a great deal of
information about the allowed values of the critical exponents at 2D filling and
wetting. Indeed, it is worthwhile developing the consequences of these
relations assuming only their validity together with the critical exponent 
relations derived earlier from standard scaling theory. Firstly, from the first
relation (\ref{law0}), it necessarily follows that 
\begin{equation}
\beta_w=\hat{\beta_s}=\frac{\zeta}{1-\zeta}
\label{Cons1}
\end{equation}
in agreement with the heuristic scaling theory. It is important to realise 
that this identification 
does $\it{not}$ depend on the specific values of the 
critical exponents pertinent to the SFL regime since the divergence of
$l_\pi(\theta)$ as $\theta \to 0$ is determined by the critical exponent 
$\hat{\beta_s}$ (\ref{hat}) rather than by $\beta_s$. This is because the covariance 
relations are expressed in terms of the angles $\theta$ and $\alpha$ rather
than the scaling fields $t$ and $t'$. Using the derived exponent relations for 
filling we can now deduce the values of the other critical exponents in the 
FFL regime 
\begin{equation}
2-\alpha_w=\frac{1-2\zeta}{1-\zeta},\Delta_w=\frac{1}{1-\zeta},\psi_w=\zeta
\label{Fset}
\end{equation}
which are all universal, determined only by $\zeta$. The second covariance law 
for the PDF's contains even more information. In
terms of the scaling functions this reads
\begin{equation}
\Lambda_w(x)=\Lambda_{\pi}^{SFL}(x)\equiv \Lambda(x)
\label{cons2}
\end{equation}
which clearly indicates that the connection with the SFL regime is fundamental and 
not a merely fortuitous coincidence of critical exponent values. The identity
of the scaling functions now has a truly remarkable consequence. To see this 
note that it necessarily follows that the SDE exponents $\gamma_w$ 
and $\gamma^{SFL}$ have the same value. Recalling the general critical exponent relations
(\ref{SDE2}),(\ref{SDEW}) for these leads to the identification
\begin{equation}
1/\beta_w=2(1/\zeta-1)-1/\beta_s
\label{Amazing!}
\end{equation}
and using the above value for $\beta_w$ we find
\begin{equation}
\beta_s=\frac{\zeta}{1-\zeta}
\label{Ivealways}
\end{equation}
which re-derives the random-walk predictions for the critical wetting SFL
regime. Thus the covariance relations severely restrict the allowed 
values of the critical exponents for both 2D filling and 2D critical wetting.

Next we turn our attention to the free-energy covariance law. Taking the derivative of 
(\ref{law2}) w.r.t. $\alpha$ we find
\begin{equation}
l_w=-\frac{\tau'(\theta-\alpha)}{2\Sigma}
\label{cons4}
\end{equation}
and setting $\alpha=0$ we arrive at a novel result relating the planar
interfacial height to the point tension: 
\begin{equation}
l_\pi(\theta)=-\frac{\tau'(\theta)}{2\Sigma}
\label{newresult}
\end{equation}
valid for interfacial models with short-ranged forces (or in the asymptotic
SFL critical regime). This relation has a number of consequences. First, 
equating the power-law 
critical singularities on either side yields  
\begin{equation}
-\frac{2\beta_s}{2-\alpha_s}=\frac{2(2-\alpha_l)}{2-\alpha_s}-1
\label{cons5}
 \end{equation}
which reduces to
\begin{equation}
\alpha_l=\alpha_s+\nu_{\parallel}
\label{ourIR}
\end{equation}
thus deriving the conjectured Indekeu-Robledo exponent relation. We emphasise that in this
manipulation we have only used the general exponent relation (\ref{Rush}) 
without having to introduce any results specific to $d=2$. This is strongly 
suggestive that a generalisation of (\ref{newresult}), with possibly different numerical
pre-factors, may well exist for SFL regime critical wetting in higher 
dimensional systems. Secondly, beyond simple power-law singularities we can now
see that the logarithmic divergence of the point tension for purely thermal 
systems found by ALU is, in fact, necessary in order that the interfacial
 height diverges as $l_\pi\sim\theta^{-1}$.

 The values of all the other critical exponents for filling and wetting now follow from 
standard exponent relations. The connection between them can be summarised by 
\begin{equation}
\beta_w=\beta_s,\alpha_w=\alpha_l,\Delta_w=\nu_{\parallel}
\label{relate}
\end{equation}
where the LHS and RHS refer to the FFL and SFL regimes respectively.

Finally, for completeness, we remark that for pure systems it has been shown \cite{ourJphysCM}
that for the marginal case $p=1$, $q=2$ corresponding to the FFL/FMF boundary, the
covariance laws for the interfacial height and PDF relate the behaviour at
filling to the WFL/MF regime of wetting. The wedge free-energy can be
easily calculated for this case and is similar to (\ref{3}) but has a numerical
pre-factor
$c=2+(1+8\Sigma b)^{1/2}$ in front of each logarithm. However
we do not discuss the possible connection with
the point tension because we are not confident that for such long-ranged forces
$\tau$ is a well-defined quantity. Whilst an expression for $\tau$ can be
found for such systems using the PDF identification, and is in accord with the
covariance law, we have not been able to extract $\tau$ using another,
independent, method. We feel such a check is necessary since, as shown by ALU,
even for short-ranged forces, the convolution definition of $\tau$ can lead to
different results. Moreover for $p=1$ it is not obvious that $\tau$ can be
extracted using a generalisation of the ALU correlation length identification. 
It may even be that the covariance relation between the
wedge-free energy and the point tension can be forwarded as a suitable
 definition of the point tension for interfacial models with this marginal
interaction. This would certainly be consistent with the Indekeu-Robledo conjecture
for the point tension singularity.

\section{Scaling and covariance for the local compressibility}

Our treatment so far has concentrated on critical singularities occurring at bulk
two-phase coexistence $h=0$. Given the precise connection between 2D filling
and wetting occurring at coexistence it is natural to enquire whether this
extends to quantities defined for $h>0$. Away from two-phase coexistence,
however, any possible relation between 2D filling and wetting is certainly 
subtler than that occurring for $h=0$ because the pertinent gap exponents 
$\Delta_w$ and $\Delta$ are different. Thus the divergences of 
$l_w\sim h^{-\psi_w}$ and $l_{\pi}\sim h^{-\psi}$ along the respective filling
and wetting critical isotherms are quite different and preclude a law of type
(\ref{law0}). Similarly there can be no simple generalisation of the free-energy
relation (\ref{law2}) because the point tension is only defined for $h=0$. This
suggests that we first look for covariance relations between response functions
for FFL filling and SFL critical wetting describing infinitesimal deviations from
bulk coexistence. Such relations, should they exist, will also be notable
because response functions are generally related to integrals over two-point
functions which would suggest that these too satisfy covariance relations.

 Before we calculate the scaling expressions for PDF and local compressibility 
for 2D filling we recall some pertinent results known for critical wetting. 

\subsection{SFL critical wetting}

Differentiating the scaling ansatz for the profile $\rho(z)$ w.r.t. $h$
immediately implies that in the scaling limit of the SFL regime, and up to an unimportant 
non-universal pre-factor $D$, the local compressibility
$\chi_{\pi}\equiv\partial \rho(z)/\partial h$ evaluated at bulk-coexistence
($h=0$) has the form \cite{Parry}
\begin{equation}
\chi_{\pi}(z;\theta)=D\theta^{-2\Delta/(2-\alpha_s)} X_{\pi}^{SFL}(z/l_{\pi}(\theta))
\label{chiSFL}
\end{equation}
where $X_\pi(x)$ is a scaling function describing the universal position
dependence. This is independent of the range of the forces and is specified 
by the dimension and type of disorder only. Again we
emphasise this is valid in the asymptotic scaling limit 
$\theta\to 0$, $z\to\infty$
with $z/l_{\pi}$ arbitrary. The SDE is controlled by the
same exponent as the density profile so that
\begin{equation}
X_{\pi}^{SFL}(x)\sim x^{\gamma^{SFL}}
\label{SDCHI}
\end{equation}
as $x\to 0$. By adopting the convention that the pre-factor is unity, the scale
of $X_{\pi}$ is fixed and we can 
regard the scaling function as universal. In effective Hamiltonian theory the behaviour
of the compressibility is directly related to that of the PDF since from
(\ref{density}) we have
\begin{equation}
\chi_{\pi}(z)=(\rho_l-\rho_v)\int_{0}^{z} d l\frac{\partial P_{\pi}(l)}{\partial h}
\label{defchi}
\end{equation}
The scaling form of the PDF, density profile and local compressibility
 emerges naturally from the interfacial model if we use the same
boundary condition (\ref{bound1}) but retain the $hl$ term in the Hamiltonian.
As shown by several authors, for $h>0$ the ground-state wave function is an 
Airy function \cite{Vall,AS,LipPRB} implying
\begin{equation}
P_{\pi}(l;\theta, h)\propto \mbox{Ai}^2 \Big(h^{1/3}l-\theta^2h^{-2/3}W_{\pi}(h\theta^{-3})\Big)
\label{FullPDF}
\end{equation}
where, for the sake of clarity, we have dropped non-universal metric factors
and written the field dependence in terms of $\theta\sim t'$. Taking into
account the $h$ dependence coming from the free-energy scaling function 
$W_{\pi}(x)$ and the normalisation constant, it is straightforward to show that
$\chi_{\pi}(z;\theta)$ scales according to the prediction (\ref{chiSFL}) with a 
universal scaling function \cite{Parry}
\begin{equation}
X_{\pi}^{SFL}(x)=(x+\frac{1}{2}x^2)e^{-x}
\label{Xp2}
\end{equation}
Notice that the SDE behaviour of this function is in accord with the general 
requirement(\ref{SDCHI}) although a different power-law determines the 
algebraic correction to the asymptotic exponential decay.

\subsection{FFL filling}

 In zero field the mid-point local compressibility, 
$\chi_w(z)\equiv\partial \rho_w(z)/\partial h$ , should also show scaling behaviour
analogous to the behaviour occurring at SFL regime critical wetting. Following 
our treatment above it follows from the profile equation (\ref{rhof}) that for $z
\to\infty$, $l_w\to\infty$ with fixed $z/l_w$,
\begin{equation} 
\chi_w(z;\theta,\alpha)=\tilde{D}(\theta-\alpha)^{-\Delta_w}X_w(z/l_w(\theta,\alpha))
\label{chif}
\end{equation}
where $X_w(x)$ is the appropriate scaling function whose SDE is described by
the critical exponent $\gamma_w$. Thus we expect
\begin{equation}
X_w(x)\sim x^{\gamma_w}
\label{etaw}
\end{equation} 
and by again adopting the convention that the critical amplitude is exactly unity 
we can fix the scale of the universal function $X_w(x)$. 

The behaviour of the mid-point PDF at filling can be easily calculated using the transfer
matrix result (\ref{p}) and after a little algebra we obtain
\begin{equation}
P_w(l;\theta-\alpha,h)\propto e^{-2\Sigma(\theta-\alpha)l-hl^2/\alpha}
\label{Gaus}
\end{equation}
which is considerably simpler then the planar result. Note that the $h$ dependence
of this scaling function has a simple geometrical meaning since the term 
$l^2/\alpha$ corresponds precisely to the area of the filled region of the
2D wedge. From (\ref{Gaus}) it is easy to check our earlier prediction that,
along the critical filling isotherm ($T=T_{\mathrm{fill}}$,$h\to 0$), the interfacial height
diverges as $l_w\sim h^{-\zeta}$. We find 
\begin{equation}
l_w\sim \big(\frac{\alpha}{\pi h}\big)^{1/2}
\label{lfh}
\end{equation}
in perfect agreement with the expected behaviour for purely thermal disorder
($\zeta=1/2$). It is also immediate from (\ref{Gaus}) that the roughness $\xi_{\perp}\sim l_w$
along the critical isotherm. We mention in passing here that this behaviour contrasts with
the singularities occurring at $\it{complete}$ filling corresponding to $h\to 0$
for $T>T_{\mathrm{fill}}$. For this transition it is apparent from (\ref{Gaus}) that the
critical behaviour is not fluctuation-dominated and $l_w\sim h^{-1}$ whilst
$\xi_{\perp}\sim h^{-1/2}$. The divergence of $l_w$ agrees with
predictions based solely on thermodynamic and mean-field arguments which
remain valid because the fluctuation effects at complete filling are small even
for the present two-dimensional system with short-ranged forces. Returning to
the local compressibility we can readily 
calculate the desired expression for the
zero-field local compressibility using the PDF (\ref{Gaus}). Taking care to account for the $h$ dependence of
the normalisation factor we find that the local compressibility is precisely of
the form (\ref{chif}) with $\Delta_w=2$ and a scaling function
\begin{equation}
X_{w}(x)=(x+\frac{1}{2}x^2)e^{-x}
\label{Xw2}
\end{equation}
which is identical to that derived for SFL regime critical wetting. Thus we
write  
\begin{equation}
X_w(x)=X_{\pi}^{SFL}(x)\equiv X(x)
\label{finallaw}
\end{equation}

In terms of the full angle dependence the scaling local compressibilities 
satisfy the simple covariance relationship
\begin{equation}
\frac{\chi_w(z;\theta,\alpha)}{\chi_{\pi}(z;\theta-\alpha)}=\frac{\theta-\alpha}{\alpha}
\label{lawchif2}
\end{equation}
valid in the asymptotic critical regime $\theta\to\alpha$. The identity of 
the local compressibility scaling functions for filling and SFL
wetting is the main result of this section. Whilst we have only demonstrated
this for pure systems we expect this is also valid for
systems filling and wetting in other 2D systems provided that $\zeta\ge1/2$. Support for
this conjecture comes from the derived values of the critical exponent
relations. In particular in the limit $z/l_{\pi}\to 0$ and $z/l_w\to 0$ the
SDE's for the local compressibilities are identical and hence their ratio must be
independent of $z$. Moreover the values of the exponents $\Delta_w$ and
$2\Delta/(2-\alpha_s)$ are always such that the ratio is proportional to
$\theta-\alpha$ independent of $\zeta$. It therefore seems highly likely that 
the covariance relation (\ref{lawchif2}) is also valid for impure systems and
can be checked in future studies. 

\section{Wedge Covariance for pure systems revisited}

Our discussion of the implications of the covariance relations in \S 5 was
simply based on comparison with the predictions of standard scaling theory for 
critical exponent relations and the short-distance expansions. With a little 
more input concerning the properties of the interfacial Hamiltonian model we can also use the
wedge-covariance relations to re-derive the results of the transfer
matrix studies without explicit calculation. The following discussion is
restricted to thermal disorder although a generalisation to the impure systems
may well be possible using the replica trick method.

In terms of the relative interfacial height $\tilde{l}-\alpha\vert x \vert$
we can write the wedge Hamiltonian as a perturbation from the planar model
\begin{equation}
H_w[\tilde{l}]=H[\tilde{l}]+\delta H_w [\tilde{l}]
\label{perturb}
\end{equation}
where
\begin{equation}
\delta H [\tilde{l}]=-2\Sigma \alpha\int dx \delta(x)\tilde{l}(x)
\label{stress}
\end{equation}
and we have ignored $\tilde{l}$-independent terms and the boundary conditions
concerning the end-points $ \tilde{l_e}$ which do not matter in the
thermodynamic limit. Because the planar interfacial Hamiltonian and the wedge
perturbation are both local it immediately follows that, up to a trivial
normalisation constant, the mid-point height PDF for pure systems satisfies
\begin{equation}
P_w(l)\propto P_{\pi}(l)e^{2\Sigma\alpha l}
\label{localward}
\end{equation}
which is equivalent to (\ref{p}). Wedge covariance of the PDF at bulk 
coexistence implies
\begin{equation}
P_{\pi}(l;\theta-\alpha)\propto P_{\pi}(l;\theta)e^{2\Sigma\alpha l}
\label{covone}
\end{equation}
and setting $\alpha=\theta$  it follows that
\begin{equation}
P_{\pi}(l;\theta)\propto P_{\pi}(l;0)e^{-2\Sigma\theta l}
\label{covtwo}
\end{equation}
where the position dependence of $P_{\pi}(l;0)$ is determined solely by the
SDE. Using the known critical exponent relation for $\gamma^{SFL}$ we conclude 
that within the SFL
regime the PDF is necessarily of the form
\begin{equation}
P_{\pi}(l;\theta)\propto l^{2/\zeta-1/\beta_s -3}e^{-2\Sigma\theta l}
\label{covthree}
\end{equation}
From this it follows that $\hat{\beta_s}=1$ and since
$\hat{\beta_s}=\zeta/(1-\zeta)$ the only value
of $\zeta$ consistent with the wedge covariance hypothesis in pure systems is
\begin{equation}
\zeta=\frac{1}{2}
\label{quantize}
\end{equation}
implying that $\beta_s=1$. Similarly the universal scaling function for the PDF in the SFL must be simply
\begin{equation}
\Lambda(x)=e^{-x}
\label{qaunttwo}
\end{equation}

Now consider the PDF off coexistence and note that the exponential boost factor
{\it does not} depend on the bulk field $h$. It follows that in the asymptotic
critical regime, the scaling functions for
filling and wetting satisfy the simple quotient relation
\begin{equation}
\frac{P_w(l;\theta,\alpha,h)}{P_w(l;\theta,\alpha,0)}\propto\frac{P_{\pi}(l;\alpha,h)}{P_{\pi}(l;\alpha,0)}
\label{quo}
\end{equation}
where the constant of proportionality is trivially determined from the
normalisation conditions on each PDF. Keeping the value of $\zeta$ arbitrary
for the moment it follows that to first order in $h$ the quotient has the
expansion
\begin{equation}
\frac{P_w(l;\theta,\alpha,h)}{P_w(l;\theta,\alpha,0)}=1+\frac{K_{\zeta
}h}{\alpha^2}(l^{1/\zeta}-\langle l^{1/\zeta}\rangle)+\dots
\label{pen}
\end{equation}
where $K_{\zeta}$ is a pure number and $\langle l^{1/\zeta}\rangle$
denotes the appropriate moment of the interfacial height evaluated at $h=0$. 
The various terms in this expansion arise for the following reasons: (1) the 
quotient must tend to unity as $h\to 0$;
(2) the expansion must be linear in $h$ because $P_{\pi}(l;\alpha,h)$ is a
non-singular function of $h$ for $\alpha>0$;
(3) the power-law dependence $l^{1/\zeta}$ is necessary since along 
the filling critical isotherm ($\alpha=\theta$, $h\to 0$) scaling demands that
 $l_w\sim h^{-\zeta}$;
(4) the dependence on $\alpha$ follows from conditions (2), (3) given the
anticipated scaling form of PDF $P_{\pi}(l;\theta,h)$;
(5) the additive term involving $\langle l^{1/\zeta}\rangle$ arises 
from the normalisation condition on each PDF.
 
 From (\ref{pen}) we immediately observe that the universal scaling functions 
$\Lambda(x)$ and $X(x)$ for FFL filling and SFL wetting must be related 
according to
\begin{equation}
X'(x)\propto \Lambda(x)(1-c_{\zeta}x^{1/\zeta})
\label{soon}
\end{equation}
where
\begin{equation}
c_{\zeta}=\frac{(\int_{0}^{\infty} s \Lambda(s)
ds)^{1/\zeta}}{\int_{0}^{\infty} s^{1/\zeta} \Lambda(s) ds}
\label{last}
\end{equation}
and the constant of proportionality in (\ref{soon}) is
trivially fixed by the condition that $X(x)\sim x^{1/\zeta-1}$ for small x. 
At this point we can substitute the appropriate expressions $\Lambda(x)=e^{-x}$ and $\zeta=1/2$
for pure systems and integrate to find
\begin{equation} 
X(x)=(x+\frac{1}{2}x^2)e^{-x}
\label{twyt}
\end{equation}
in agreement with the transfer matrix calculation. The differential equation
(\ref{soon}) is also an effective method of calculating the covariant scaling function
for the local compressibility at the FFL/FMF (or equivalently WFL/MF) 
borderline with $p=1$,$q=2$ for which the PDF function $\Lambda(x)$ 
has a non-trivial SDE
\cite{ourJphysCM}. Further work is required to
see if similar approaches can determine the allowed values of $\zeta$,
$\Lambda(x)$ and $X(x)$ for disordered systems without using the full transfer
matrix formulism.
 
\section{Discussion}

 In this paper we have investigated fluctuation effects occurring at 2D filling
 and the fundamental connection with the 
SFL regime of critical wetting. Our study has revealed that in addition to the
known covariance of the interfacial height and PDF
\cite{ourJphysCM,ourJphysCM2}, the wedge free-energy and
mid-point local susceptibility are also related to the point tension and 
local susceptibility at wetting. These
relations are extremely restrictive and determine the allowed values of the 
critical exponents at FFL filling $\it{and}$ SFL wetting without further 
assumptions 
other than those of standard scaling theory. Moreover if wedge covariance is
combined with knowledge of how the wedge
Hamiltonian is perturbed from the planar interfacial model, then very specific
predictions for the value of the wandering exponent, critical exponents and
scaling functions can be obtained. Thus wedge covariance appears to 
play a similar role to the principle of conformal invariance for 2D bulk critical
phenomena in that it yields predictions over and above those of scaling and
scale invariance. Wedge covariance also bring new insights into the 
nature of the SFL regime wetting transition itself. In particular we have shown that the 
covariance relation for the wedge free-energy provides a means of deriving 
the conjectured Indekeu-Robledo relation for the critical singularity of the 
point tension and also explains its logarithmic divergence for pure systems.
Other aspects of our work that have not been presented before include the 
derivation of an expression for the point tension for impure systems and the
development of a scaling theory and derivation of critical exponent relations 
for 2D filling transitions.

It is hoped that the present work motivates the further study of filling
transitions in both two and three dimensional, pure and impure systems. Within
the framework of the present, small $\alpha$ interfacial model, future work 
could include discussion of two-point functions. These may well exhibit some
covariance properties as
suggested by the behaviour of the local compressibility. Non-linear functional 
renormalisation group analyses, both approximate and exact decimation \cite{Forgacs}
type, 
would also shed light on the likely expanded space of the SFL regime fixed 
point. Studies of fluctuation effects at filling using other models such as 
the full drumhead interfacial Hamiltonian or lattice Ising model, for which
only limited results are known \cite{ourPRL3}, would also be very welcome. It
would also be interesting to see if the formal statistical mechanical theory
of fluids at interfaces, with its powerful sum-rule and correlation function
hierarchies \cite{Row,Henderson1,Henderson2}, can be applied to the problem. Perturbative
expansions of the many-body Hamiltonian analogous to (\ref{perturb}) and
(\ref{stress}) may prove revealing. Staying within the framework of effective
Hamiltonian theory it would be highly informative if the nature of both 3D
wedge and cone filling were understood beyond the case of purely thermal
disorder. It is likely that this would also shed more light on the nature of
the SFL regime in higher dimensions about which very little is currently known.
Nevertheless the simple heuristic 
picture of how the filling of a 2D wedge
manages to precisely mimic the behaviour of the SFL regime,
does generalise rather naturally to the cone geometry which similarly enforces the same
qualitative type of conic interfacial configuration that determines large
 deviations in the one-point function at wetting. It is tempting to speculate that the
Indekeu-Robledo critical exponent relation may also be intimately tied to a
possible covariance for three dimensional wedge free-energy although much 
further work is required to quantify this.

AOP and AJW are very grateful to Dr.\ Carlos Rasc\'on for discussions and
collaboration on related work. AJW and MG thank the EPSRC for financial
support.


\section*{References}
\begin{thebibliography}{30}
\bibitem{Finn} P.\ Concus, R.\ Finn, Proc.\ Nat.\ Acad.\ Sci.\
U.S.A., {\bf 63}, 292 (1969). See also R.\ Finn, {\it Equilibrium
Capillary Surfaces}, Springer, New York, 1986, and, P.\ Concus, R.\ Finn, Phys. Fluids {\bf 10}, 39 (1998).

\bibitem{Pomeau} Y.\ Pomeau, J.\ Colloid Interface Sci. {\bf 113}, 5 (1986).
\bibitem{Hauge} E.\ H.\ Hauge, Phys.\ Rev.\ A {\bf 46}, 4994 (1992).
\bibitem{MFT} K.\ Rejmer, S.\ Dietrich and M.\ Napi\'{o}rkowski, Phys.\
Rev.\ E {\bf 60} 4027, (1999).
\bibitem{ourPRL1}  A.\ O.\ Parry, C.\ Rasc\'{o}n and A.\ J.\ Wood, Phys.\ Rev.\ Lett.\ {\bf 83}, 5535 (1999).
\bibitem{ourJphysCM} A.\ O.\ Parry, A.\ J.\ Wood. and C.\ Rasc\'{o}n, J.\ Phys. Cond.\ Matt.\ {\bf 12}, 7671 (2000).
\bibitem{ourPRL2} A.\ O.\ Parry, C.\ Rasc\'{o}n and A.\ J.\ Wood. Phys.\ Rev.\ Lett.\ {\bf 85}, 345 (2000).
\bibitem{ourJPCMlett}  A.\ J.\ Wood and A.\ O.\ Parry, J.\ Phys.\ A:\
Math.\ Gen.\
{\bf 34}, L5 (2001).
\bibitem{ourJphysCM2} A.\ O.\ Parry, A.\ J.\ Wood. and C.\ Rasc\'{o}n, J.\ Phys. Cond.\ Matt.\ {\bf
13}, 4591 (2001).
\bibitem{Nap} A.\ Bednorz and Napi\'{o}rkowski, J.\ Phys.\ A:\ Math.\ Gen.\
{\bf 33}, L353 (2001).
\bibitem{ourPRL3} A.\ O.\ Parry, A.\ J.\ Wood, Enrico Carlon and A.\
Drzewi\'nski. Phys.\ Rev.\ Lett.\ . To appear, (2001).
\bibitem{General} For general reviews of wetting see for example
M. Schick,  (1990) in {\it Liquids at interfaces: Les Houches,  Session
XLVIII} ed.\ J.\ Charvolin,  J.\ F.\ Joanny and J.\ Zinn-Justin and 
S. Dietrich in {\it Phase Transitions and Critical
Phenomena},  ed. C.\ Domb and J.L.\ Lebowitz, Vol.\ {\bf 12}, p.\ 1
(Academic Press, London, 1988). 
\bibitem{Forgacs} The fluctuation theory of wetting, including
disordered systems, is discussed at length in G.\ Forgacs, R.\ Lipowsky 
and Th.\ M.\ Nieuwenhuizen in {\it Phase Transitions and Critical
Phenomena} , {\bf 14},ed. C.\ Domb and J.\ Lebowitz, New
York: Academic (1991). 
\bibitem{Lipowsky} R.\ Lipowsky,  J.\ Phys.\ A.\ : Math.\ and Gen.\ {\bf
18}, L585 (1985).
\bibitem{Fisher} M.\ E.\ Fisher, J.\ Chem.\ Soc.\ Faraday Trans.\ 2 {\bf
82}, 1589 (1986).
\bibitem{LandF} R.\ Lipowsky and M.\ E.\ Fisher, Phys. Rev. B {\bf
36}, 2126 (1987).
\bibitem{Wuttke} J.\ Wuttke and R.\ Lipowsky, Phys.\ Rev.\ B, {\bf 44},
13042 (1991).
\bibitem{BPZ} see for example J.\ L.\ Cardy in {\it Phase Transitions
and Critical Phenomena} , {\bf 11}, ed. C.\ Domb and J.\ Lebowitz, New
York: Academic (1988). 
\bibitem{IR} J.\ O.\ Indekeu and A.\ Robledo, Phys.\ Rev.\ E.\, {\bf 47}, 4607 (1993).
\bibitem{I} J.\ O.\ Indekeu, Int.\ J.\ of Mod.\ Phys.\ B, {\bf 8},
309 (1993).
\bibitem{ABU} D.\ B.\ Abraham, F.\ T.\ Latr\'emoli\`ere and P.\ J.\
Upton, Phys.\ Rev.\ Lett.\ {\bf 71}, 404 (1993).
\bibitem{Row} J.\ S.\ Rowlinson and B.\ Widom, {\it Molecular Theory
of Capillarity}, Clarendon Press, Oxford (1982).
\bibitem{Abraham} D.\ B.\ Abraham, Phys.\ Rev.\ Lett.\ {\bf 44}, 1165 (1980).
\bibitem{Burkhardt1} T.\ W.\ Burkhardt, J. Phys. A {\bf 14}, L63 (1981). 
\bibitem{Chalker} J.\ T.\ Chalker, J. Phys. A {\bf 14}, 2431 (1981).
\bibitem{Kardar} M.\ Kardar, Phys.\ Rev.\ Lett.\ {\bf 55}, 2235 (1985).
\bibitem{Burkhardt2} T.\ W.\ Burkhardt, Phys. Rev. B {\bf 40}, 6987 (1989).
\bibitem{Parry} A.\ O.\ Parry, J. Phys. A {\bf 24}, 1335 (1991). 
\bibitem{Parry2} A.\ O.\ Parry, J. Phys. A {\bf 24}, L699 (1991). 
\bibitem{Parry3} A.\ O.\ Parry and R.\ Evans, Mol.\ Phys.\ {\bf 78}, 1527 (1993).
\bibitem{DB} C.\ Bauer and S.\ Dietrich, Phys.\ Rev.\ E {\bf 62}, 2428 (2000).
\bibitem{Steph} T.\ Pompe and S.\ Herminghaus, Phys.\ Rev.\ Lett.\ {\bf
85}, 1930 (2000).
\bibitem{Binder} E.\ V.\ Albano, K.\ Binder and D.\ W.\ Heermann, Surf.\
Sci.\ {\bf 223}, 151 (1989).
\bibitem{Parry and Evans1} A.\ O.\ Parry and R.\ Evans, Phys.\ Rev.\
Lett.\ {\bf 64}, 439 (1990).
\bibitem{Parry and Evans2} A.\ O.\ Parry and R.\ Evans, Physica A {\bf
181}, 250 (1992).
\bibitem{AU} D.\ B.\ Abraham, N.\ M.\ $\breve{\mbox{S}}$vraki\'c and P.\ J.\ Upton, Phys.\
Rev.\ Lett.\ {\bf 68}, 423 (1992).
\bibitem{Dux} P.\ M.\ Duxbury and A.\ C.\ Orrick, Phys.\ Rev.\ B {\bf
39}, 2944 (1989).
\bibitem{Adam} A.\ Lipowski, Phys.\ Rev.\ E {\bf 58}, R1 (1998).
\bibitem{Vall} M.\ Vallade and J.\ Lajzerowicz, J.\ Phys.\ (Paris) {\bf
42}, 1505 (1981).
\bibitem{AS} D.\ B.\ Abraham and E.\ R.\ Smith, Phys.\ Rev.\ B {\bf 26},
1480 (1982), see also D.\ B.\ Abraham in {\it Phase Transitions and Critical
Phenomena - Volume 10}, edited by C.\ Domb and J.\ L.\ Lebowitz,
Academic Press, London (1988). 
\bibitem{LipPRB} R. Lipowsky, Phys.\ Rev.\ E {\bf 32}, 1731 (1985).
\bibitem{Henderson1} For a review see Henderson J.\ R.\ in {\it
Fundamentals of inhomogeneous fluids} edited by Douglas Henderson, New
York, Marcel Dekker (1992).
\bibitem{Henderson2} Henderson J.\ R.\, Mol.\ Phys.\ {\bf 98}, 677 (2000).
\end{thebibliography}
\end{document}